\documentclass[useAMS,usenatbib]{mn2e}
\usepackage{graphicx}
\usepackage{epstopdf}
\usepackage[figuresright]{rotating}
\usepackage{subfigure,color}
\usepackage{longtable}
\usepackage{xspace}

\newcommand{\ii}{\'\i}
\newcommand{\ion}[2]{{\textrm{#1}}\,{\textrm{\sc #2}}}

\title{Interaction effects on  galaxy pairs  with Gemini/GMOS- I: Electron density}

\author[Krabbe et al.]
{A.~C.~Krabbe$^{1}$\thanks{E-mail:angela.krabbe@gmail.com}, 
D.~A.~Rosa$^1$, O.~L.~Dors Jr.$^1$, M.~G.~Pastoriza$^2$,  C.~Winge$^3$, \newauthor 
G.~F.~H\"agele$^{4,5}$, M.~V.~Cardaci$^{4,5}$ and I.~Rodrigues$^1$\\
$^1$ Universidade do Vale do Para\'iba, Av. Shishima Hifumi, 2911, Cep
12244-000, S\~ao Jos\'e dos Campos, SP, Brazil\\
$^2$ Instituto de F\ii sica, Universidade Federal do Rio Grande do Sul, Av.~Bento Gon\c{c}alves, 9500,  Cep 91359-050, Porto Alegre, RS, Brazil\\
$^3$ Gemini Observatory, c/o AURA Inc., Casilla 603, La Serena, Chile \\
$^4$ Instituto de Astrof\'isica de La Plata (CONICET La Plata--UNLP), Argentina. \\
$^5$ Facultad de Ciencias Astron\'omicas y Geof\'isicas, Universidad Nacional de La Plata, Paseo del Bosque s/n, 1900 La Plata, Argentina \\
}
\begin{document}

\date{Accepted -. Received -.}

\pagerange{\pageref{firstpage}--\pageref{lastpage}} \pubyear{2012}

\maketitle

\label{firstpage}

%________________________________________________________________

\begin{abstract}
We present an observational study about the impacts  of the interactions in the electron density 
of \ion{H}{ii} regions located in  7 systems of interacting galaxies.
The data consist of long-slit spectra  in the range 4400-7300 \AA{}, obtained with the Gemini Multi-Object Spectrograph at Gemini South (GMOS).
The electron density was determined using the ratio of emission lines [S{\scriptsize\,II}]$\lambda$6716/$\lambda$6731. Our results indicate that the
 electron density estimates  obtained of  \ion{H}{ii} regions from our sample of interacting galaxies are systematically higher than those derived for isolated galaxies. The mean  electron   density values  of interacting galaxies are in the range of $N_{\rm e}=24-532$\,$ \rm cm^{-3}$, while those obtained for isolated galaxies are in the range of $N_{\rm e}=40-137\: \rm cm^{-3}$.  Comparing the observed emission lines with predictions of photoionization models,   we verified that almost  all the   \ion{H}{ii} regions of the galaxies  AM\,1054A, AM\,2058B, and  AM\,2306B, have emission lines excited by shock gas. For the remaining galaxies, only  few
\ion{H}{ii} regions has emission lines excited by shocks, such as in AM\,2322B (1 point), and AM\,2322A (4 points). No correlation is obtained  between the presence of shocks and electron densities. Indeed,  the highest electron density values found
 in our sample do not belong to the objects with  gas shock excitation.  We emphasize the importance of considering theses quantities  especially when the metallicity is derived for these types of systems.
\end{abstract}

\section{Introduction}

The study of physical processes involved in galaxy collisions and mergers in the local universe is fundamental to understand 
the formation and evolution of these objects, providing important
constraints in  simulations of  the universe at large scale. 

In particular, the chemical abundance is highly modified in interacting/merging galaxies.
\citet{kewley10} presented a systematic investigation about
metallicity gradients in  close pairs of galaxies. These authors 
determined the oxygen abundance (generally used as  a tracer of the
metallicity $Z$) along the disk of eight  galaxies in close pairs and found 
metallicity gradients shallower than the ones in isolated
galaxies. Similar results have been reached by \citet{krabbe11,krabbe08}, who 
built spatial profiles of oxygen abundance of the gaseous phase of the
 galaxy pairs AM\,2306-721 
and AM\,2322-2821. This flattening in the oxygen abundance gradient reflects  the effects of gas redistribution along the galaxy disk due to metal-poor inflow  
of gas  from outskirts of the  centre of interacting galaxies \citep{rupke10}.

The large-scale gas motion created by the interaction induces high star formation rate and
galactic-scale outflows  \citep{veilleux05}, producing shock excitation in  star-forming regions, such as reported in recent  studies of Luminous Infrared
Galaxies by \citet{soto12} and \citet{rich12, rich11}.  In particular, \citet{rich11},  through  integral field spectroscopic  data
of the Luminous Infrared Galaxies IC\,1623 and NGC\,3256,  showed that broad line profiles
are often associated with gas shock excitation in \ion{H}{ii} regions located in mergers. 
Similar results were also found by  \citet{newman12} for the clumpy star-forming galaxy 
ZC\,406690 (see also \citealt{soto12a}). These authors  pointed out that
the  broad emission likely originates from large-scale outflows with high 
 mass rates from individual star-forming regions. 
The changes in galaxies that experience an encounter seem to have a relation with the separation among the objects interacting, such
as showed by \citet{scudder12}. These authors, using spectroscopy data of 
a large  sample of objects with a close companion taken from Sloan Digital Sky Survey Data
Release, found that the metallicity gradient and the star formation rate (SFR) are correlated with the separation of the galaxy pairs analysed,
in the sense that  the  gradients are flatter and the SFR are higher at
smaller separations. 

Despite recent efforts to probe the properties of interacting galaxies, the
electron density of star-forming regions have been poorly determined in these
systems, as well as its correlation with other quantities (e.g. $Z$, SFR).
In galaxy  disks of interacting galaxies, where gas motions and 
 gas excited by shock are present,  high electron density is expected and, can be used  as a signature
of the presence of these  motions and shocks. In fact, 
\citet{puech06}, in a study about galaxy interaction, mapped electron densities in six distant galaxies (z\,$\sim$\,0.55)
 and found that
 the highest electron density values observed could be associated to the
collision between molecular clouds of the interstellar medium and gas
inflow/outflow events.
These authors derived electron density values lower than 400 $\rm
cm^{-3}$, typical of classical \ion{H}{ii} regions \citep{copetti00,castaneda92}.  
However, \citet{puech06} used as a sensor the
[\ion{O}{ii}]$\lambda$3729/$\lambda$3727  ratio, which  underestimates the electron
density in relation to 
determinations via other line ratios \citep{copetti02}.
Most of oxygen determinations of the gas phase  in interacting galaxies (e.g. \citealt{scudder12,rich12,rich11,krabbe11,kewley10,krabbe08})
are based on theoretical models that consider low electron density values of  10-200 $\rm cm^{-3}$
\citep{krabbe11,dors11, kewley02, dopita00}.
If the electron density values considerably differ of those considered in
 the models, the oxygen abundance estimations will be doubtful. 
In fact, \citet{oey93} showed that systematic variations in the nebular density introduce 
significant uncertainties into the  abundances obtained using methods based on strong emission lines. They found that
differences between 10 and 200 $\rm cm^{-3}$, a  typical range 
 for the electron density derived in giant \ion{H}{ii} regions 
(e.g. \citealt{copetti00,castaneda92,kennicutt84,odell84}), reflect
variations  up to 0.5 dex in oxygen abundances, mainly
for the high metallicity regime.  These variations can increase even more
when higher electron density values, such as the ones 
found in  star-forming clumps (e.g.\ 300-1800 $\rm cm^{-3}$,
\citealt{newman12}), are considered in abundance determinations. 
 
In this paper, we used long-slit spectroscopic data of a sample of seven pair galaxies to verify the effects 
of the interaction on the electron density in these systems. This work is organized as follows. In Section~\ref{obser_data}, we
summarize the observations and data reduction. In Section~\ref{determinations}, 
the method to compute the electron density 
is described. Results and discussion are
presented in Sections~\ref{results} and \ref{discussion}, respectively.
The conclusions of the outcomes are given in Section~\ref{conc}.

\begin{table*}
\caption{Galaxy sample.}
\label{sample}
\begin{tabular}{llllrrl}
\hline
\noalign{\smallskip}
ID         & Morphology   &$\alpha$(2000) & $\delta$(2000)                                          & $cz\,(\rm km/s)$ & $m_{\rm B}$\,(\rm mag)  & Others names \\
\hline
\noalign{\smallskip}
AM\,1054-325  & Sm    [2]       & $10^{\rm{h}}56^{\rm{m}}58\fs2$      & $-33^{\rm{h}}09^{\rm{m}}52\fs0$       &  3\,788 [10]    & 14.55 [2]  & ESO 376-IG 027     \\
              & Sa    [5]       & 10 ~57 ~04.2s                       & $-$33 ~09 ~21.0                       &  3\,850 [5]    & 15.41 [8]  & ESO 376- G 028     \\
%%%%%%%%%%%%%%%%%%%%%%%%%%%%%%%%%%%%%%%%%%%%%%%%%%%%%%%%%%%%%%%%%%%%%%%%%%%%%%%%%%%%%%%%%%%%%%%%%%%%%%%%%%%%%%%%%%%%%%%%%%%%%%%%%%%%%%%%%%%%%%%%%%%%%%%%%%%%%%%%%
AM\,1219-430  & Sm    [6]       & 12 ~21 ~57.3                        & $-$43 ~20 ~05.0                       &  6\,957 [3]     & 14.30 [7]  & ESO 267-IG 041       \\
              & S?    [6]       & 12 ~22 ~04.0                        & $-$43 ~20 ~21.0                       &  6\,879 [3]     &   -~~~~        & FAIRALL 0157          \\
%%%%%%%%%%%%%%%%%%%%%%%%%%%%%%%%%%%%%%%%%%%%%%%%%%%%%%%%%%%%%%%%%%%%%%%%%%%%%%%%%%%%%%%%%%%%%%%%%%%%%%%%%%%%%%%%%%%%%%%%%%%%%%%%%%%%%%%%%%%%%%%%%%%%%%%%%%%%%%%%%
AM\,1256-433  & E     [3]       & 12 ~58 ~50.9                        & $-$43 ~52 ~30.0                       &  9\,215 [3]     & 14.75 [8]  & ESO 269-IG 022 NED01  \\
              & E     [3]       & 12 ~58 ~50.6                        & $-$43 ~52 ~53.0                       &  9\,183 [3]     & 16.17 [8]  & ESO 269-IG 022 NED02 \\
              & SBC   [3]       & 12 ~58 ~57.6                        & $-$43 ~50 ~11.0                       &  9\,014 [3]     & 16.41 [1]  & ESO 269-IG 023 NED01  \\
%%%%%%%%%%%%%%%%%%%%%%%%%%%%%%%%%%%%%%%%%%%%%%%%%%%%%%%%%%%%%%%%%%%%%%%%%%%%%%%%%%%%%%%%%%%%%%%%%%%%%%%%%%%%%%%%%%%%%%%%%%%%%%%%%%%%%%%%%%%%%%%%%%%%%%%%%%%%%%%%%
AM\,2058-381  & Sbc   [6]       &21 ~01 ~39.1                         & $-$38 ~04 ~59.0                       &  12\,383 [3]    &14.91 [1]   & ESO 341- G 030         \\
              & ?               &21 ~01 ~39.9                         & $-$38 ~05 ~53.0                       &  12\,460 [3]    &16.24 [1]   & ESO 341- G 030 NOTES01  \\
%%%%%%%%%%%%%%%%%%%%%%%%%%%%%%%%%%%%%%%%%%%%%%%%%%%%%%%%%%%%%%%%%%%%%%%%%%%%%%%%%%%%%%%%%%%%%%%%%%%%%%%%%%%%%%%%%%%%%%%%%%%%%%%%%%%%%%%%%%%%%%%%%%%%%%%%%%%%%%%%%
AM\,2229-735  & SO?   [3]       &22 ~33 ~43.7                         & $-$73 ~40 ~47.0                       &  17\,535 [3]    &15.98 [1]   & AM 2229-735 NED01    \\
              & ?               &22 ~33 ~48.3                         & $-$73 ~40 ~56.0                       & 17\,342 [3]     &17.36 [1]   & AM 2229-735 NED02   \\
%%%%%%%%%%%%%%%%%%%%%%%%%%%%%%%%%%%%%%%%%%%%%%%%%%%%%%%%%%%%%%%%%%%%%%%%%%%%%%%%%%%%%%%%%%%%%%%%%%%%%%%%%%%%%%%%%%%%%%%%%%%%%%%%%%%%%%%%%%%%%%%%%%%%%%%%%%%%%%%%%
AM\,2306-721  & SAB(r)c         &23 ~09 ~39.3                         & $-$71 ~01 ~34.0                       &  8\,919 [4]     &14.07 [1]   & ESO 077- G 003  \\
              &  ?              &23 ~09 ~44.5                         & $-$72 ~00 ~04.0                       &  8\,669 [4]     &14.47 [1]   & ESO 077-IG 004  \\ 
AM\,2322-821  &  SA(r)c         &23 ~26 ~27.6                         & $-$81 ~54 ~42.0                       &  3\,680 [3]     &13.35 [1]   & ESO 012- G 001, NGC 7637  \\
              &  ?              &23 ~25 ~55.4                         & $-$81 ~52 ~41.0                       &  3\,376 [4]     &15.41 [1]   & ESO 012- G 001 NOTES01     \\
\hline
\noalign{\smallskip}
\end{tabular}
\begin{minipage}[c]{1\textwidth}
{\it References:} [1] \citet{ferreiro04}; [2] \citet{weilbacher00}; [3] \citet{donzelli97}; [4] \citet{krabbe11};
 [5] \citet{lauberts82}; [6] \citet{paturel03}; [7] \citet{devaucouleurs91}; [8] \citet{lauberts89}; [9] \citet{huchra12}; [10] \citet{jone09} \\
{\it Conventions:} $\alpha$, $\delta$: equatorial coordinates\\
\end{minipage}
\end{table*}

%##################################################################################################################
\
\section{Observations and Data Reduction}
\label{obser_data}

We have selected several systems from \citet{ferreiro04}
to study the effects of the kinematics, stellar population, gradient
abundances, and electron densities of  interacting galaxies.  The first results
of this programme were presented for AM2306-721 \citep{krabbe08}
and AM2322-821 \citep{krabbe11}. 
Table \ref{sample} summarizes the main characteristics of the
 systems:  identification, morphology, position, radial velocity, apparent B magnitude, and 
 other designations.

Long-slit spectroscopic data were obtained on May, June, and July 2006 and 2007;  and  July 2008 with the 
Gemini Multi-Object Spectrograph (GMOS-S) attached to the 8\,m Gemini South 
telescope, Chile,  as part of  the poor weather programs GS-2006A-DD-6,  GS-2007A-Q-76, and  GS-2008A-Q-206. 
Spectra in the range 4400-7300 \AA{} were acquired with  the
B600 grating, and 1$\arcsec$ slit  width,  assuming a compromise between spectral
resolution (5.5\,\AA{}), spectral coverage, and slit losses (due to the Image Quality = ANY constraint).
The frames were binned on-chip by 4 and 2 pixels in the spatial and spectra
directions, respectively, resulting in a spatial scale of  0.288\,$\arcsec$px$^{-1}$,
and dispersion of 0.9 \AA{} px$^{-1}$. 

\begin{figure*}
\centering
\includegraphics[width=0.47\textwidth]{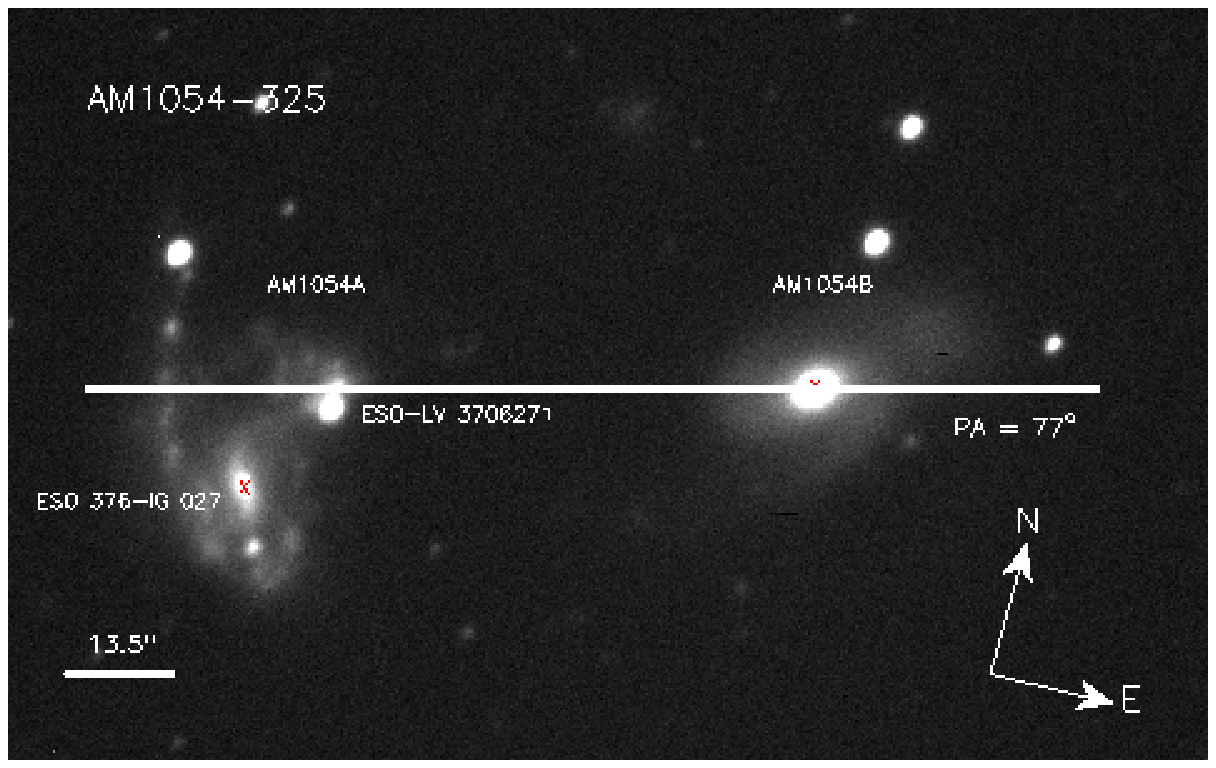}
\includegraphics[width=0.47\textwidth]{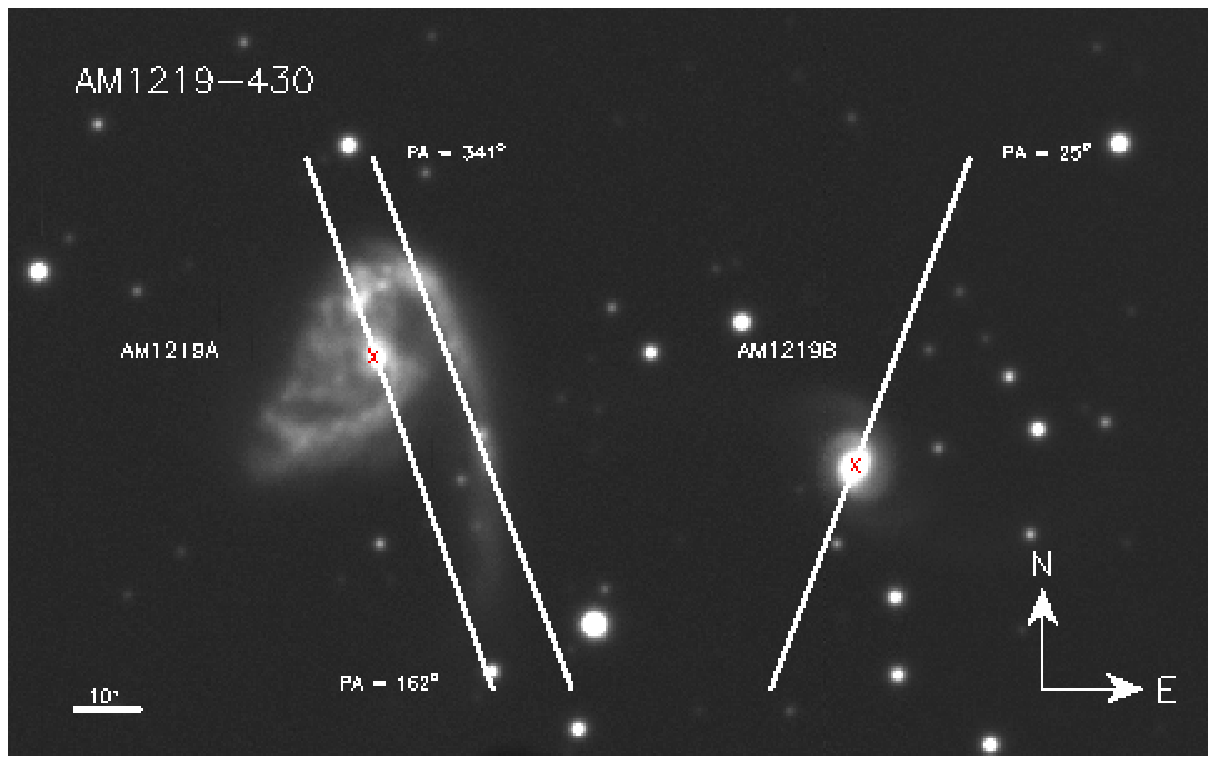}\\
\includegraphics[width=0.47\textwidth]{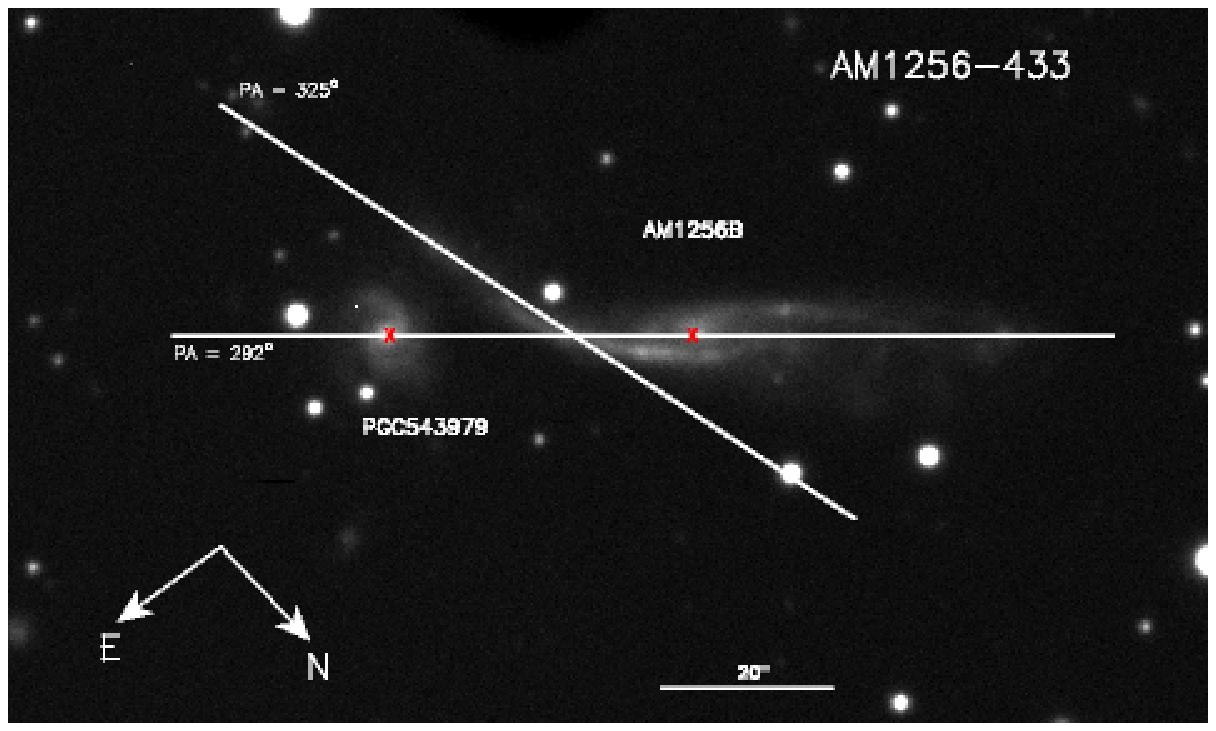}
\includegraphics[width=0.47\textwidth]{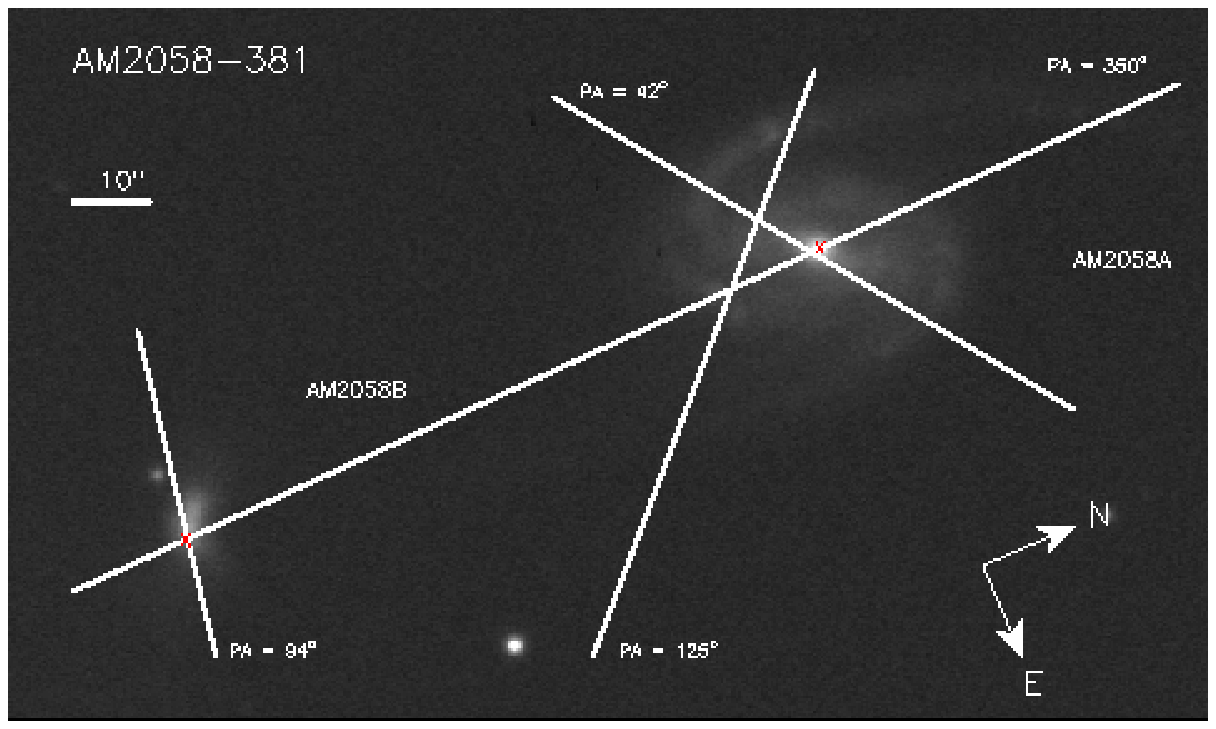}\\
\includegraphics[width=0.47\textwidth]{am2229_janeiro2013.eps}
\includegraphics[width=0.47\textwidth]{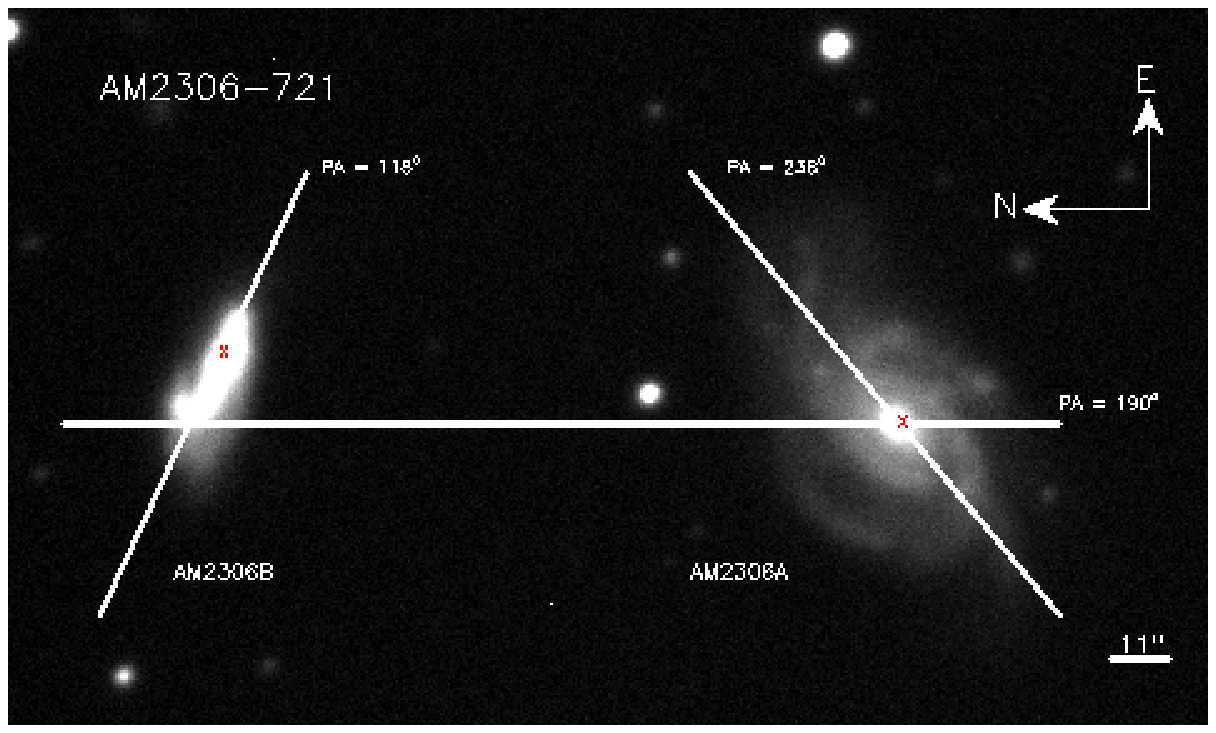}\\
\includegraphics[width=0.47\textwidth]{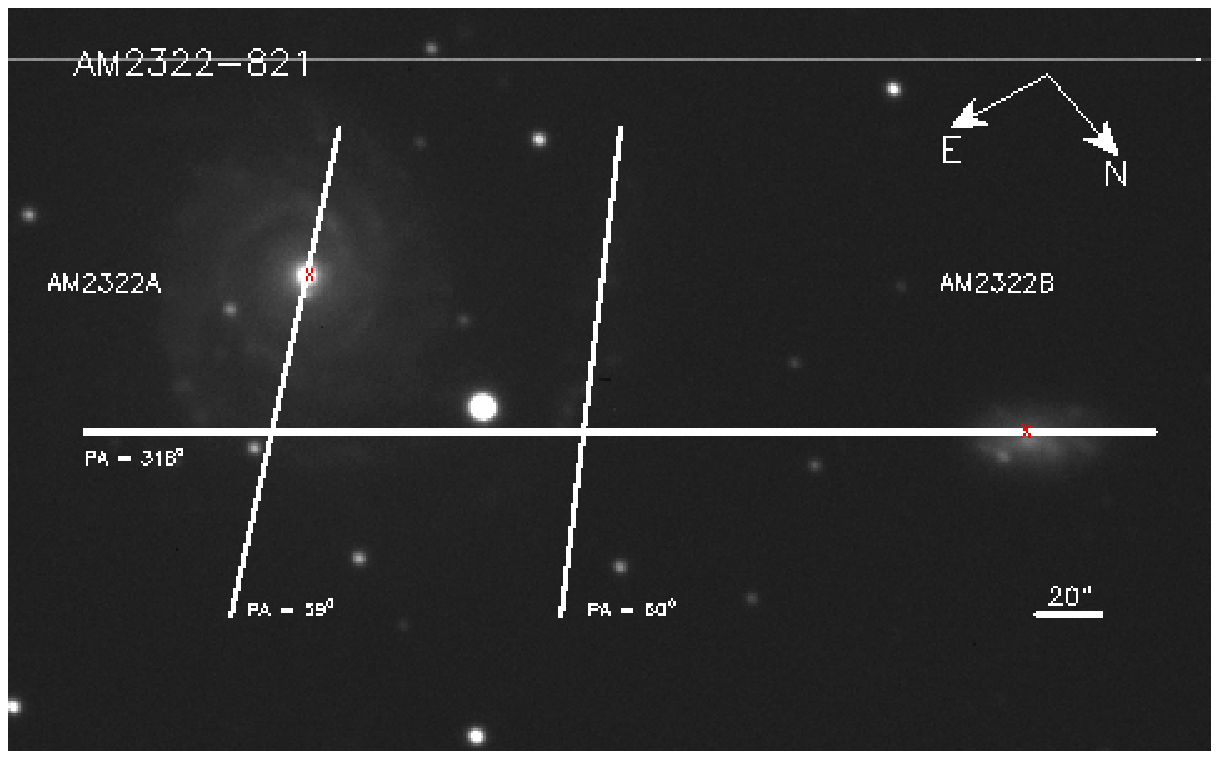}
\caption{The slit positions for each system are shown superimposed 
on the GMOS-S r$\arcmin$ acquisition image.}
\label{fendas}
\end{figure*}

Spectra were taken at different position angles on the sky,
with the goal of observing the nucleus and the brightest regions
of the galaxies.  The exposure time on each single frame was limited to 700
seconds to minimize the effects of cosmic rays, with multiple
frames being obtained for each slit position to achieve a suitable
signal. The slit positions for each system are shown in
Fig.~\ref{fendas}, superimposed on the GMOS-S
r$\arcmin$ acquisition images.  Table  \ref{tab1} gives the journal of
observations. 
Conditions during  the observing runs were not photometric, with thin cirrus
and image quality in the range 0.6$\arcsec$ to 1.7$\arcsec$ (as measured from
stars in the acquisition images taken just prior to the spectroscopic
observations).

The spectroscopic data reduction was carried out  using 
the {\sc gemini.gmos} package and generic {\sc IRAF}\footnote{Image Reduction and  
Analysis Facility, distributed by NOAO, operated by AURA, Inc., under
agreement with NSF.} tasks.
We followed the standard procedure: (1) the data were bias subtracted and flat-fielded; 
(2) the wavelength calibration was established from the Cu-Ar arc frames with
typical residuals of 0.2 \AA\, and applied to the object frames; (3) the
individual spectra of same slit positions and wavelength range were averaged
with cosmic ray rejection; (4) the object frames were sky subtracted
interactively using the {\sc gsskysub} task, which uses a background sample of
off-object areas to fit a function to the  specified rows, and this fit was
then subtracted from the column of each spectra; (5) the spectra were relative
flux calibrated using observations of a flux standard star taken with the same
set up as the science observations; (6) finally, one-dimensional spectra were
extracted  from the two-dimensional spectra by  summing over four rows along
the spatial direction. Each spectrum, therefore, comprises the flux contained in
an aperture of 1$\arcsec \times 1.152 \arcsec$. 

The  intensities of the H$\beta$, [\ion{O}{iii}]$\lambda$5007, [\ion{O}{i}]$\lambda$6300,
H$\alpha$, [\ion{N}{ii}]$\lambda$6584, and [\ion{S}{ii}]$\lambda\,$6716,$\lambda$\,6731 
emission lines were measured using  a single Gaussian line profile
fitting on the spectra. We used the  
{\sc IRAF splot} routine to fit
the lines, with the associated error being given as $\sigma^{2} =\sigma_{cont}^{2} + \sigma_{line}^{2} $, where $\sigma_{cont}$ and\,
$\sigma_{line}$ are the continuum  rms and the Poisson error of the line flux, respectively. Furthermore, we considered only measurements 
whose continuum around $\lambda$ 6700 \AA\, reach a 
signal-to-noise S/N  $\geq$ 8. The emission line intensities were not corrected for the interstellar
extinction, because it  is negligible due to the small separation between the [\ion{S}{ii}]$\lambda6716$ and $\lambda6731$ emission lines.

\begin{table}
\caption{Journal of observations.}
\label{tab1}
\begin{tabular}{@{} llcll @{}}
\noalign{\smallskip}
\hline
\noalign{\smallskip}
Object      &Date   & Exposure & PA (\degr)& $\Delta\,\lambda$ (\AA) \\ 
\noalign{\smallskip}
            &       & Time~(s) &           &  \\ 
\noalign{\smallskip}
\hline
\noalign{\smallskip}
AM\,1054-325 & 2007-06-21 & 4 $\times$  600   & 77    & 4280-7130\\
AM\,1219-430 & 2007-06-06 & 4 $\times$  600   & 25    & 4280-7130\\
             & 2007-05-26 & 4 $\times$  600   & 162   & 4280-7130\\
	     & 2007-06-22 & 4 $\times$  600   & 341   & 4280-7130\\
AM\,1256-433 & 2007-07-06 & 4 $\times$  600   & 292   & 4280-7130 \\
	     & 2007-06-21 & 4 $\times$  600   & 325   & 4280-7130 \\ 
AM\,2058-381 & 2006-05-20 & 4 $\times$  600   & 42    & 4351-7213 \\ 
	     & 2007-05-26 & 4 $\times$  600   & 94    & 4351-7213 \\ 
             & 2007-05-24 & 4 $\times$  600   & 125   & 4351-7213 \\ 
             & 2007-05-30 & 4 $\times$  600   & 350   & 4351-7213 \\
AM\,2229-735 & 2006-07-20 & 6 $\times$  600   & 134   & 4390-7250 \\
	     & 2006-07-16 & 6 $\times$  600   & 161   & 4390-7250 \\
AM\,2306-721 & 2006-06-20 & 4 $\times$  600   & 118   & 4280-7130 \\
	     & 2006-06-20 & 4 $\times$  600   & 190   & 4280-7130 \\
 	     & 2006-06-20 & 4 $\times$  600   & 238   & 4280-7130 \\
AM\,2322-821 & 2006-07-01 & 3 $\times$  700   & 59    & 4280-7130\\
	     & 2008-07-27 & 6 $\times$  600   & 60    & 4280-7130\\
	     & 2006-06-30 & 6 $\times$  600   & 318   & 4280-7130\\ 
\noalign{\smallskip}	     
\hline
\noalign{\smallskip}
\end{tabular}
\end{table}

%##################################################################################################################

\begin{figure*}
\centering
\includegraphics*[angle=-90,width=\textwidth]{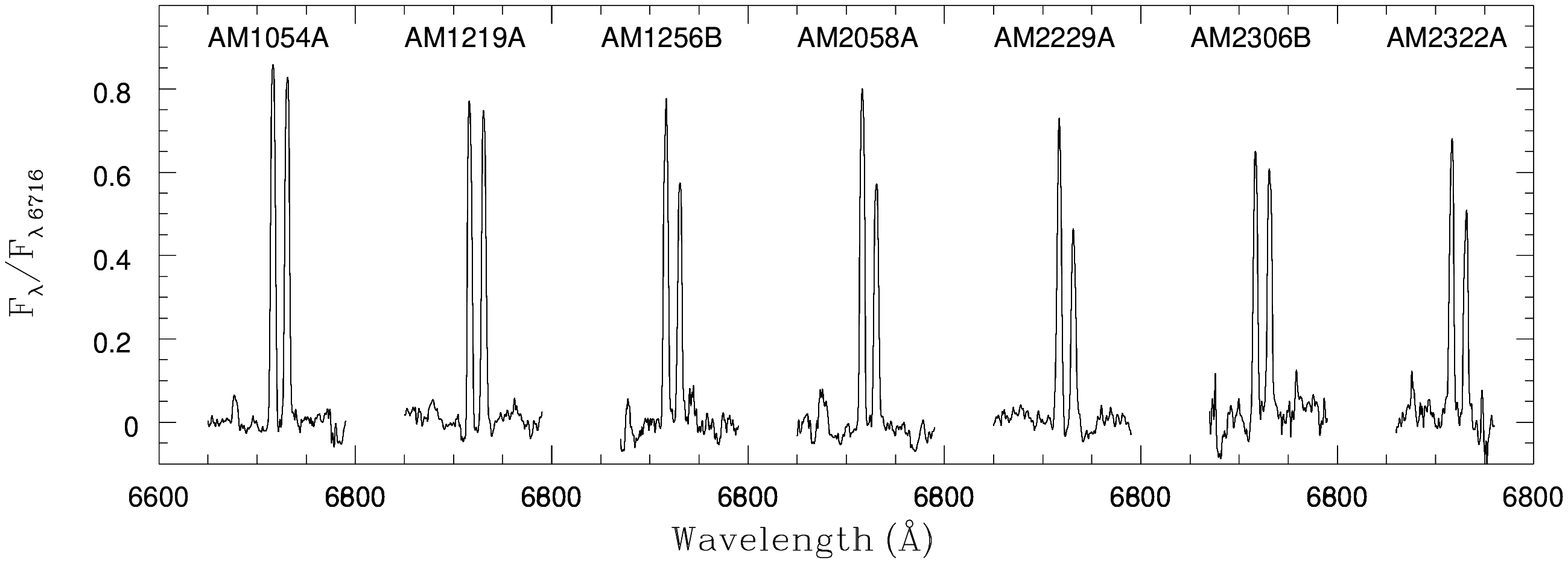}
\caption{A sample of spectra in the range of 6600 to 6800 \AA\, from areas of different galaxies. The flux scale was normalized to the peak of [\ion{S}{ii}]\,$\lambda$\,6716.}
\label{sample_sii}
\end{figure*}

\section{Determination of the electron density}
\label{determinations}

The electron density  $N_{\rm e}$ was derived from the
[\ion{S}{ii}]$\lambda\,$6716/$\lambda$\,6731 emission line  intensity ratio
by solving numerically the equilibrium equation for a 
$n$-level atom approximation using  the  {\sc temden} routine of the {\sc nebular} package of the {\sc STSDAS/IRAF},  assuming an electron 
temperature of 10\,000\,K, because  temperature sensitive
emission lines were unobservable in our sample.

The references for the collision strengths,  transition probabilities, and energy levels are \citet{Ramsbottom96}, \citet{verner87}, \citet{keenan93},
and  \citet{bowen60}. There are two main sources of errors  in the determination  of electron  densities. 
One is the dependence of the $N_{\rm e}$ on the electron temperature  $T_{\rm e}$ assumed. 
However,    this dependence is weak
in the range of temperatures usually found in galactic \ion{H}{ii} regions
\cite[e.g.][]{copetti00}. 
We adopted a mean  electron temperature  of 10\,000 K as a representative value,
because it is a typical electron temperature value for these kinds of objects
and there are no estimations for our sample.
The other main source of error is the saturation of  the line ratio for
both low and high values of  the electron density, which makes the [\ion{S}{ii}]$\lambda$\,6716/$\lambda$\,6731 ratio a reliable sensor of 
the electron density in the range of 2.45\,$<$\,log\,$N_{\rm e}$\,(cm$^{-3})$\,$<$\,3.85 \citep{stanghellini89}.

%##################################################################################################################

\section{Results}
\label{results}

Fig.~\ref{sample_sii} shows a sample of  the spectra of  some \ion{H}{ii} regions
of the galaxies  around the [\ion{S}{ii}]$\lambda$6716 and  [\ion{S}{ii}]$\lambda$6731
emission lines.  The profiles of $\log$([\ion{O}{i}]$\lambda$6300/\rm H$\alpha$),  
[\ion{S}{ii}]$\lambda$6716/$\lambda$6731 ratio, and 
$N_{\rm e}$ as a function of galactocentric radius for the galaxies are shown
in Figs. \ref{den_am1054}-\ref{den_am2322b}. 
 The intensity of  $\log$([\ion{O}{i}]$\lambda$6300/\rm H$\alpha$) was plotted
only for the apertures  for which the  electron  density determination was possible.
The galactocentric radius are not corrected by galaxy inclination.
 In Fig.\ \ref{fendas} the adopted centre of each galaxy  is marked with a red cross.
Table \ref{media}  presents  some statistics of the
[\ion{S}{ii}]\,$\lambda$\,6716/$\lambda$\,6731 ratio  and electron density
measurements, including the number  {\it N} of distinct nebular areas, the
mean, the median, the  maximum and minimum, and the standard deviation
$\sigma$. The results for each system are presented  separately.

\begin{table*}
\caption{[\ion{S}{ii}] ratio and electron density statistics.} 
\label{media}
\begin{footnotesize}
\begin{tabular}{lcccccccccccccc} 
\hline 
\noalign{\smallskip}
 &  &   \multicolumn{6}{c}{[\ion{S}{ii}] $\lambda\,6716/\lambda\,6731$} & &\multicolumn{6}{c}{$N_{{\rm e}} ~ (\mathrm{cm^{-3}})$ }    
\cr 
\cline{3-8}
\noalign{\smallskip} 
\cline{10-15}
\noalign{\smallskip} 
Objects & & N & mean & median  & max &    min &   $\sigma$ & &  N &  mean &   median & max &  min & $\sigma$ \\ 
\hline
\noalign{\smallskip} 
AM\,1054A   && 16  & 1.19  &  1.08  &1.70   & 0.97   & 0.93 &&  13 & 434  & 462 & 681 & 65 & 191  \\
AM\,1054B   &&  3  & 0.86  &  0.85  &0.92   & 0.79   & 0.07 &&   3 &1130  & 1082&1476 & 833& 324\\
AM\,1219A   && 29  & 1.12  &  1.06  &1.77   & 0.86   &0.23  &&  26 & 532  & 518 &1073 & 85 & 286 \\
AM\,1219B   &&  5  & 0.70  & 0.82   &0.92   &0.23    &0.27  &&   4 & 1408 &1294 &2189 &855 & 564\\
AM\,1256B   && 43  & 1.48  &  1.42  &2.08   & 0.99   &0.27  &&	22 & 181  & 317 & 626 &  7 & 168 \\			 
AM\,2058A   && 20  & 1.38  &  1.26  &2.22   & 0.90   &0.37  &&  13 & 376  & 318 & 911 & 33 & 263 \\
AM\,2058B   && 8   & 1.47  &  1.42  &1.80   & 1.24   &0.19  &&	 4 &  86  & 60  & 184 & 42 & 66 \\
AM\,2229A   &&33   & 1.60  &  1.59  &2.61   & 0.19   &0.59  &&	 7& 346   & 226 &686  & 28 &280 \\
AM\,2306A   &&8    & 1.41  &  1.44  &1.60   & 1.16   &0.16  &&	 5& 131   & 107 &298  & 32 &99 \\
AM\,2306B   &&15   &1.38   &  1.30  &2.88   & 0.92   & 0.47 &&	 11& 300  & 212 & 826 & 19 &273 \\
AM\,2322A   &&81   &1.41   &1.42    &1.89   & 0.85   & 0.20 &&	 41& 200  & 103 & 1121& 11 &259\\
AM\,2322B   &&23   &1.47   &1.43    &1.74   &1.35    &0.10  &&   12& 24   & 15  & 75  & 3  &23\\
\noalign{\smallskip} 
\hline
\end{tabular}
\end{footnotesize}
\end{table*}

%##################################################################################################################
\subsection{AM\,1054-325}

This system  is composed by a peculiar spiral with disturbed arms (hereafter AM\,1054A) and a spiral-like object  (hereafter AM\,1054B).   
AM\,1054A  contains very luminous  H{\scriptsize\,II} regions
along their galactic disk. As can be seen in the Fig. \ref{fendas}, AM\,1054A 
seems to have two nuclei. According to measurements obtained from \citet{weilbacher00}, the 
``main'' nucleus of this galaxy (ESO 376-IG 027)  is the reddest [(B-V)=0.52], while the other (ESO-LV 3760271) has the blue colours of a strong  
starburst  [(B-V)=0.21]. Both nuclei
names are marked   in the Fig. \ref{fendas}.
The  measured radial velocity is 3788 km/s \citep{jone09} and 3853 km/s \citep{sekiguchi93}
for ESO 376-IG 027 and ESO-LV 3760271, respectively.  Therefore, the
  small difference found  between their radial velocities  together with the perturbed morphology
of the galaxy seem to indicate that these objects are gravitationally bound.

\begin{figure*}
\centering
\includegraphics*[angle=-90,width=0.85\textwidth]{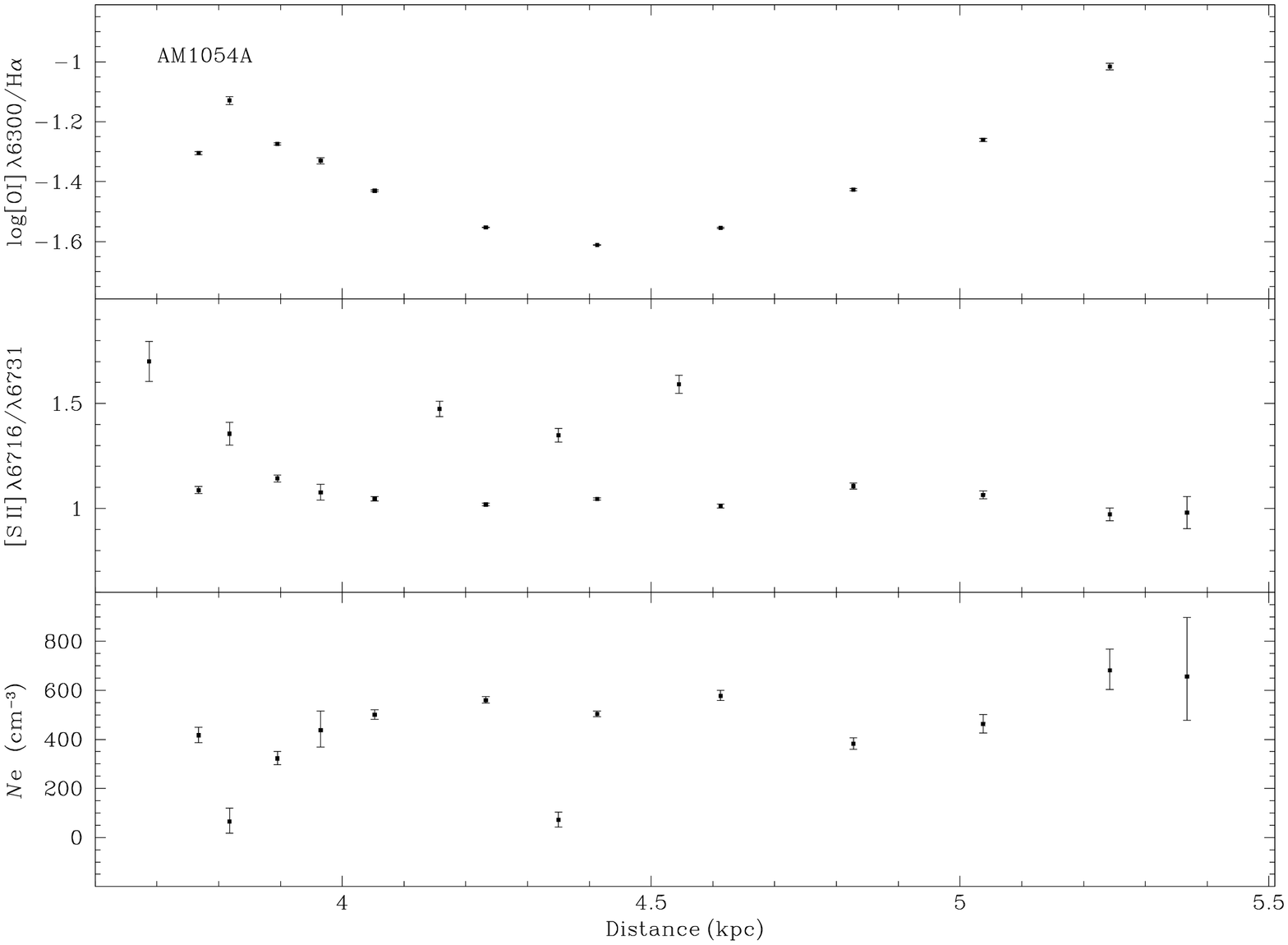}
\caption{AM\,1054-325. $\log$([\ion{O}{i} $\lambda$\,6300/\rm H$\alpha$) ratio,  
[\ion{S}{ii}] $\lambda$\,6716/$\lambda$\,6731 ratio, and 
$N_{\rm e}$ as a function of galactocentric radius for AM\,1054A.}
\label{den_am1054}
\includegraphics*[angle=-90,width=0.85\textwidth]{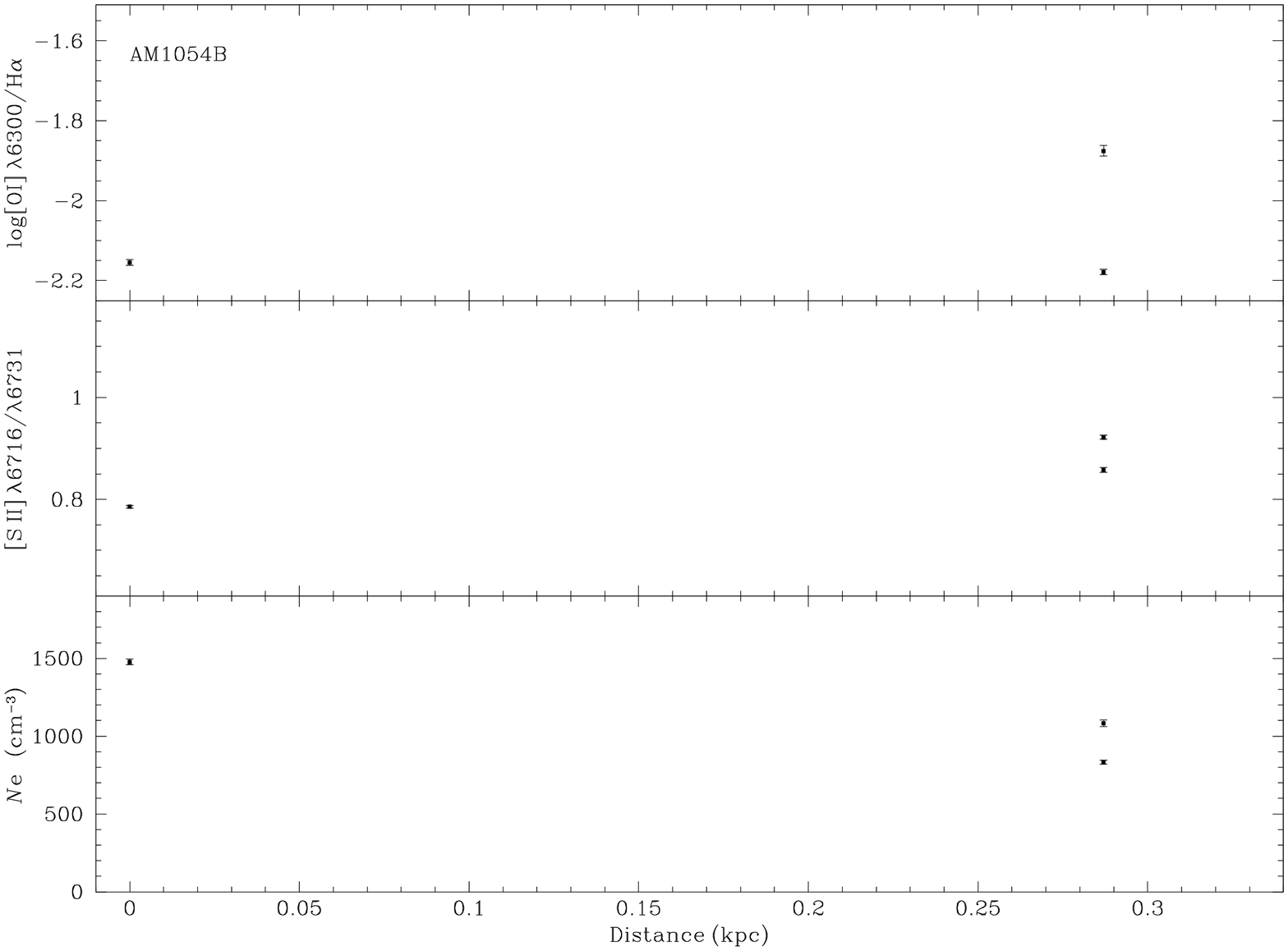}
\caption{Same as Fig. \ref{den_am1054}, but for AM\,1054B.}
\label{den_am1054b}
\end{figure*}

For AM\,1054A, the electron  density values estimated from [\ion{S}{ii}]$\lambda\,6716/\lambda\,6731$ ratio (see  Fig. \ref{den_am1054}) 
present variations  of relatively high  amplitude along the radius of the galaxy, with the minimum
  value of $N_{\rm e}$= 65 ~$\mathrm{cm^{-3}}$ and the maximum of  $N_{\rm e}$= 681 ~$\mathrm{cm^{-3}}$. We found a mean density of  $N_{\rm e}= 434 \pm 53$~$\mathrm{cm^{-3}}$.
In this galaxy, the slit position is cutting a bright star-forming region, but does not cross the
nucleus of the galaxy. For AM1054B, only  few apertures had the
[\ion{S}{ii}]$\lambda\lambda$\,6716, 6731 emission lines  with enough
  signal to be
measured. A mean density of  $N_{\rm e}= 1\,130 \pm 187 $~$\mathrm{cm^{-3}}$ was derived for this galaxy.

%##################################################################################################################

\subsection{AM\,1219-430}

This pair is composed by a disturbed spiral  (hereafter AM\,1219A) and a smaller disk galaxy (AM\,1219B).
AM\,1219A  shows a tidal tail produced by the interaction of the galaxies, with very bright \ion{H}{ii} regions. Systemic velocities of 6\,957 km\,s$^{-1}$
and 6\,879 km\,s$^{-1}$ were estimated by \citet{donzelli97} for AM\,1219A and
AM1219B, respectively.   

The distribution of electron densities exhibits variations of high amplitude
across the radius of the main galaxy in the range of $N_{\rm e}= 85 - 1073
$~$\mathrm{cm^{-3}}$. We found a mean density of $N_{\rm e}= 532 \pm 56
$~$\mathrm{cm^{-3}}$.   As in the case of AM\,1054B, only for few apertures
  of AM\,1219B the [\ion{S}{ii}]$\lambda\lambda$\,6716,6731 emission lines
  have enough signal to be measured. A mean density of $N_{\rm e}= 1\,408 \pm 282 $~$\mathrm{cm^{-3}}$
 was derived for this galaxy.  Interestingly, the $N_{\rm  e}$   increases toward the outskirt of this galaxy. This region is at the end of the spiral arm to the Northwest.

\begin{figure*}
\centering
\includegraphics*[angle=-90,width=0.85\textwidth]{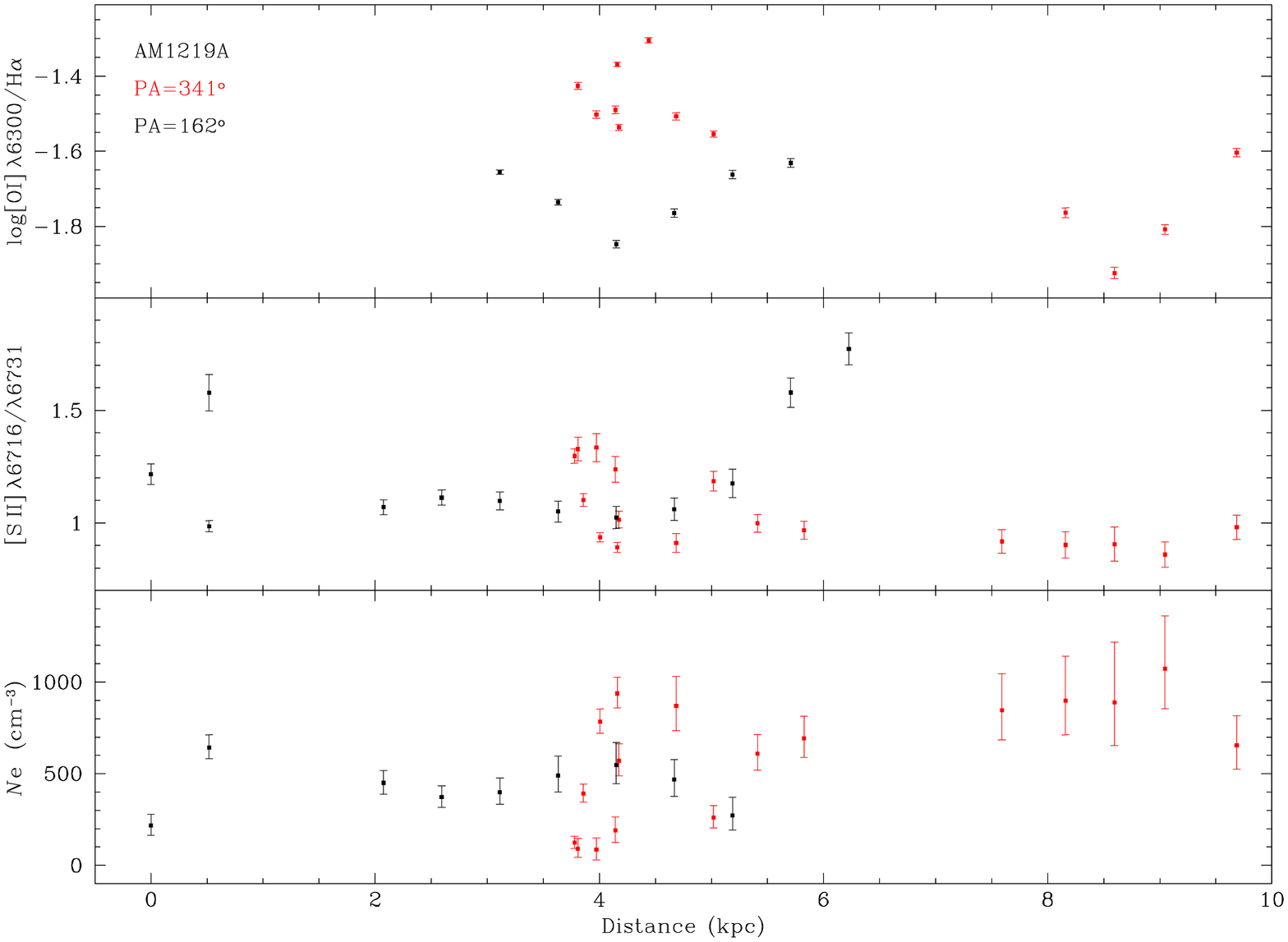}
\caption{Same as Fig. \ref{den_am1054}, but for AM\,1219A.}
\label{den_am1219a}
\includegraphics*[angle=-90,width=0.85\textwidth]{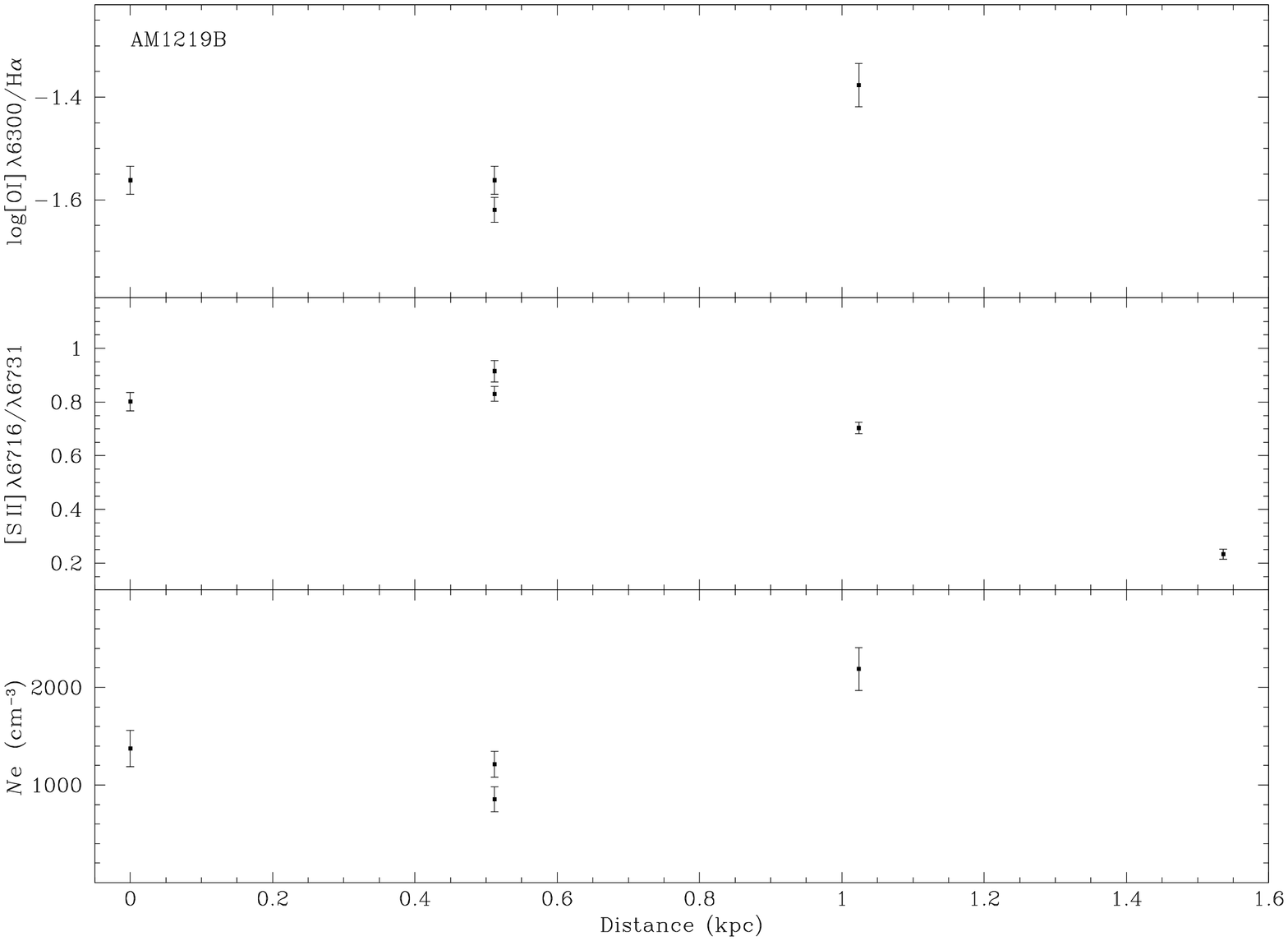}
\caption{Same as Fig. \ref{den_am1054}, but for AM\,1219B.}
\label{den_am1219b}

\end{figure*}

\subsection{AM\,1256-433}

AM\,1256-433 is a system constituted by three galaxies.  Two are elliptical with very bright nuclei, 
ESO 269-IG 022 NED01 and ESO 269-IG 022 NED02, and  one  very  disturbed spiral galaxy, ESO 269-IG 023 NED01, 
hereafter AM\,1256B. In addition, an isolated disk galaxy, ESO 269-IG 023
NED02/PGC 543979  ($\alpha=12^{\rm{h}}59^{\rm{m}}00\fs6$ and $\delta=-43^{\rm{h}}50^{\rm{m}} 23^{\rm{s}}$ J2000), 
appears in the  field of view of this system, about 30$\arcsec$  to the
Southeast of the centre of AM\,1256B. 
From our data, we obtained for this isolated galaxy a 
heliocentric velocity of 18\,896\,km\,s$^{-1}$ indicating that it does not
belong to this system, and it was incorrectly associated with AM\,1256-433 by \citet{donzelli97,ferreiro04}, and \citet{ferreiro08}.  
In Fig.~\ref{fendas}, only AM\,1256B and the isolated galaxy  ESO 269-IG 023 NED02  is shown.

\begin{figure*}
\centering
\includegraphics*[angle=-90,width=0.85\textwidth]{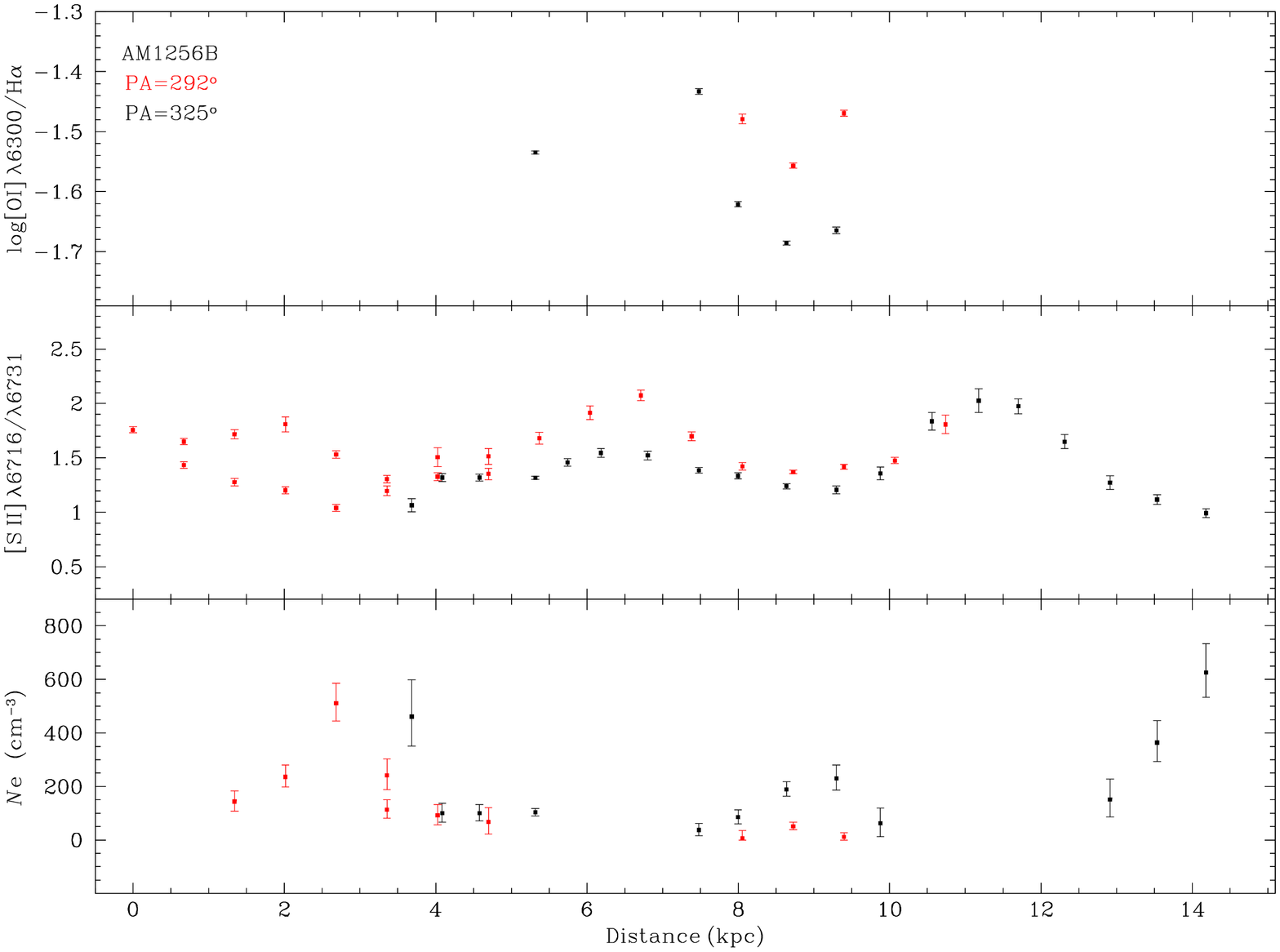}
\caption{Same as Fig. \ref{den_am1054}, but for AM\,1256B.}
\label{den_am1256}
\end{figure*}

 As can be seen in Fig. \ref{den_am1256}, some regions 
(for example at about 6 and 12 kpc from the centre of the galaxy)
present un-physically large values of the [\ion{S}{ii}]$\lambda$6716/$\lambda$6731 ratio,  above the theoretical value of 1.4,  the
value for the low density limit according to the \cite{osterbrock06} curve for this relation.
There could be some uncertainties associated with the measurements of these
sulphur emission lines, due to the placement of the continuum and deblending
of the lines, that might produce larger values of the [\ion{S}{ii}] ratio
than the expected ones. Values of the [\ion{S}{ii}] ratio larger than the 1.4
upper limit were already observed in other studies using different kinds of
instruments \cite[e.g.][]{kennicutt89,zaritsky94,lagos09,relano10,lopez13}.
As pointed out by \cite{lopez13},  the theoretical density determination also
needs to be adjusted to the sulphur atomic data and deserves to be revisited. 
From a spatial distribution study of the electron density in a sample of
\ion{H}{ii} regions in M33, these authors highlighted that when values of the
$\lambda$\,6716/$\lambda$\,6731 ratio above the 1.4 limit are obtained, it is
reasonable to assume that the electron densities are lower than 10
$\mathrm{cm^{-3}}$. They also noted that a safe way to proceed is to take
$N_{\rm e}= 100 $~$\mathrm{cm^{-3}}$, because even before reaching the 1.4 limit,
the estimation of the electron density is very uncertain.
A mean density of  $N_{\rm e}= 181 \pm 36 $~$\mathrm{cm^{-3}}$ was derived for
this galaxy.  Again,  the $N_{\rm  e}$ increases toward the outskirt of this galaxy, 
corresponding to the end of the spiral arm at Southeast.

\subsection{AM\,2058-381}
This system of galaxies is a typical M\,51 type pair. It  has a systemic velocity of ${cz}$\,=\,12\,286\,km\,s$^{-1}$ \citep{donzelli97} and consists of a 
main galaxy with two spiral arms (hereafter, AM\,2058A) and a  companion irregular galaxy (hereafter, AM\,2058B). 

\begin{figure*}
\begin{center}
\includegraphics*[angle=-90,width=0.85\textwidth]{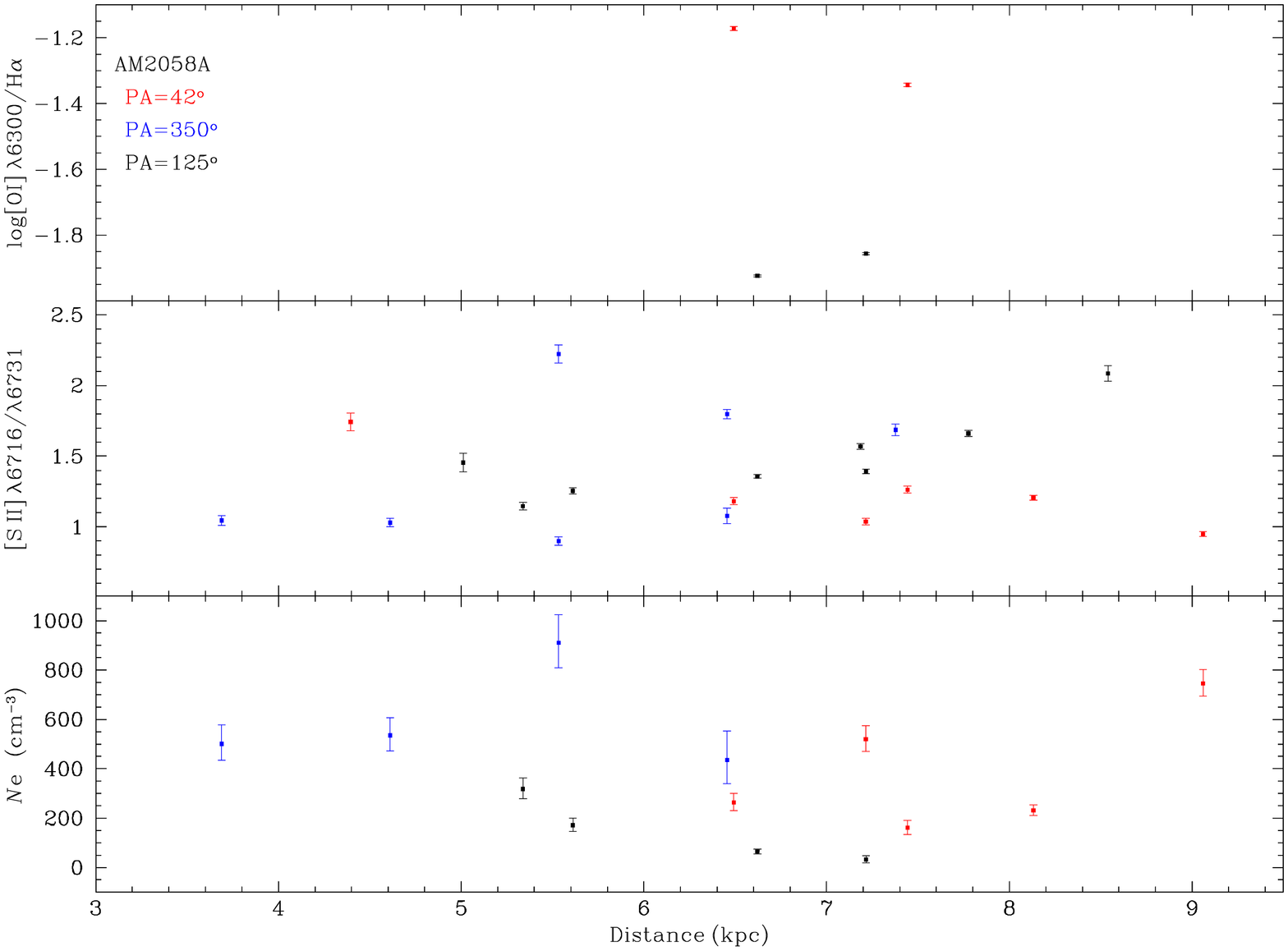}
\caption{Same as Fig. \ref{den_am1054}, but for AM\,2058A.}
\label{den_am2058a}
\includegraphics*[angle=-90,width=0.85\textwidth]{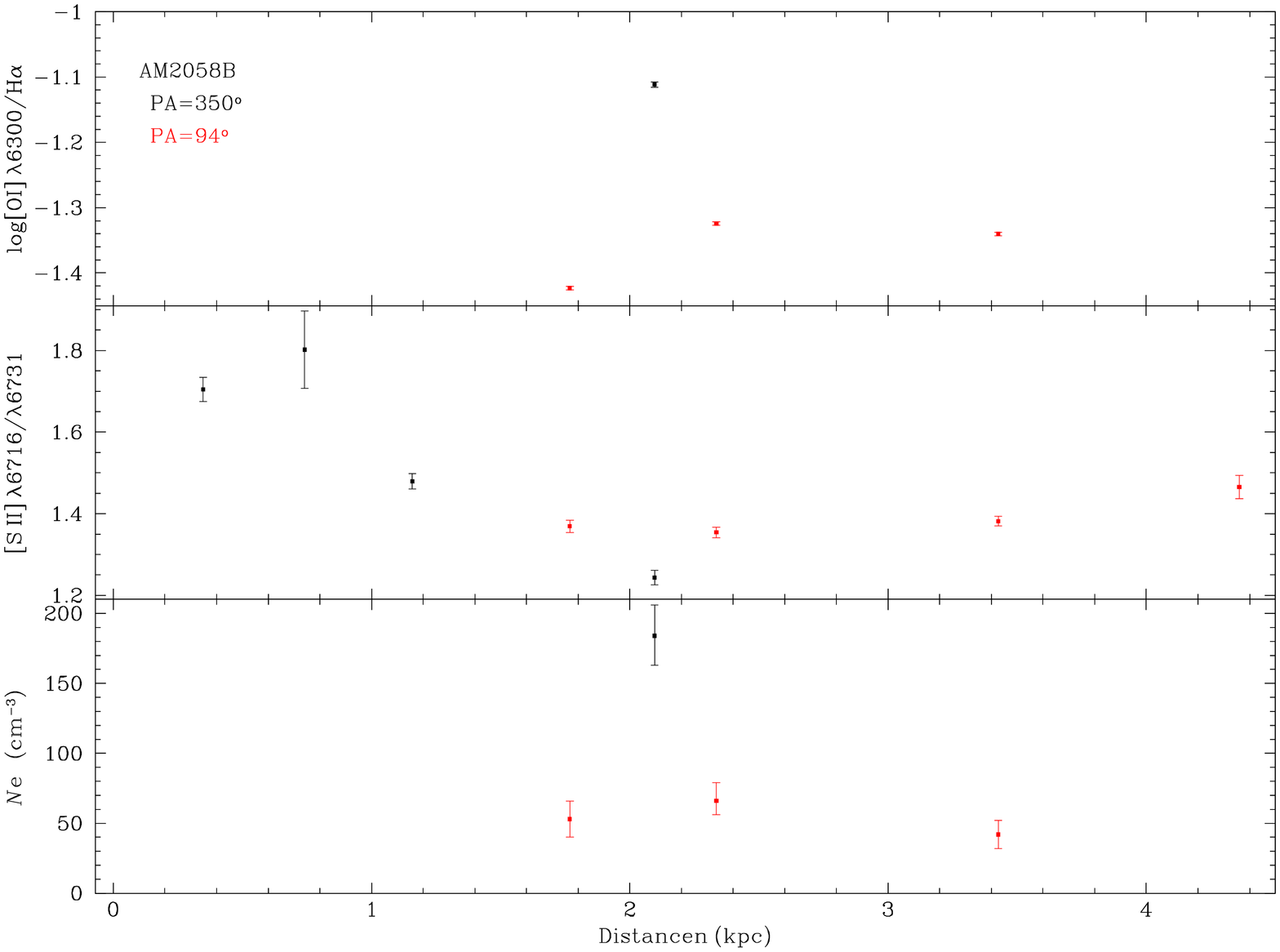}
\caption{Same as Fig. \ref{den_am1054}, but for AM\,2058B.}
\label{den_am2058b}
\end{center}
\end{figure*}

The electron densities obtained for AM\,2058A have variations across the galaxy in the range of $N_{\rm e}= 33-911  $~$\mathrm{cm^{-3}}$, and these values are not dependent upon the position.  
Due to the small radius of AM\,2058B,  only a few apertures could be extracted
for  this galaxy. The electron densities  (see Fig.~\ref{den_am2058b}) are relatively low, 
with a mean value of $N_{\rm e}= 86 \pm 33 $~$\mathrm{cm^{-3}}$, which is
compatible with  estimations for giant extragalactic \ion{H}{ii} regions 
(e.g. \citealt{castaneda92}).

\subsection{AM\,2229-735}

This  pair of galaxies consists of a main spiral galaxy strongly disturbed (hereafter AM\,2229A) and a smaller disk galaxy that could be 
connected to the main one by a bridge. AM\,2229A  has a very massive nucleus of $M= 5\times 10^{8} M_{\sun}$ \citep{ferreiro08} and very bright \ion{H}{ii}  regions. 
Only the  primary galaxy was observed.

 Most of observed regions in AM\,2229A present un-physically large values of the
\ion{S}{ii} ratio according to the \cite{osterbrock06} curve for the relation
between this ratio and the electron density.
We derived a mean electron density of $N_{\rm e}= 346 \pm 95 $~$\mathrm{cm^{-3}}$.

\begin{figure*}
\centering
\includegraphics*[angle=-90,width=0.85\textwidth]{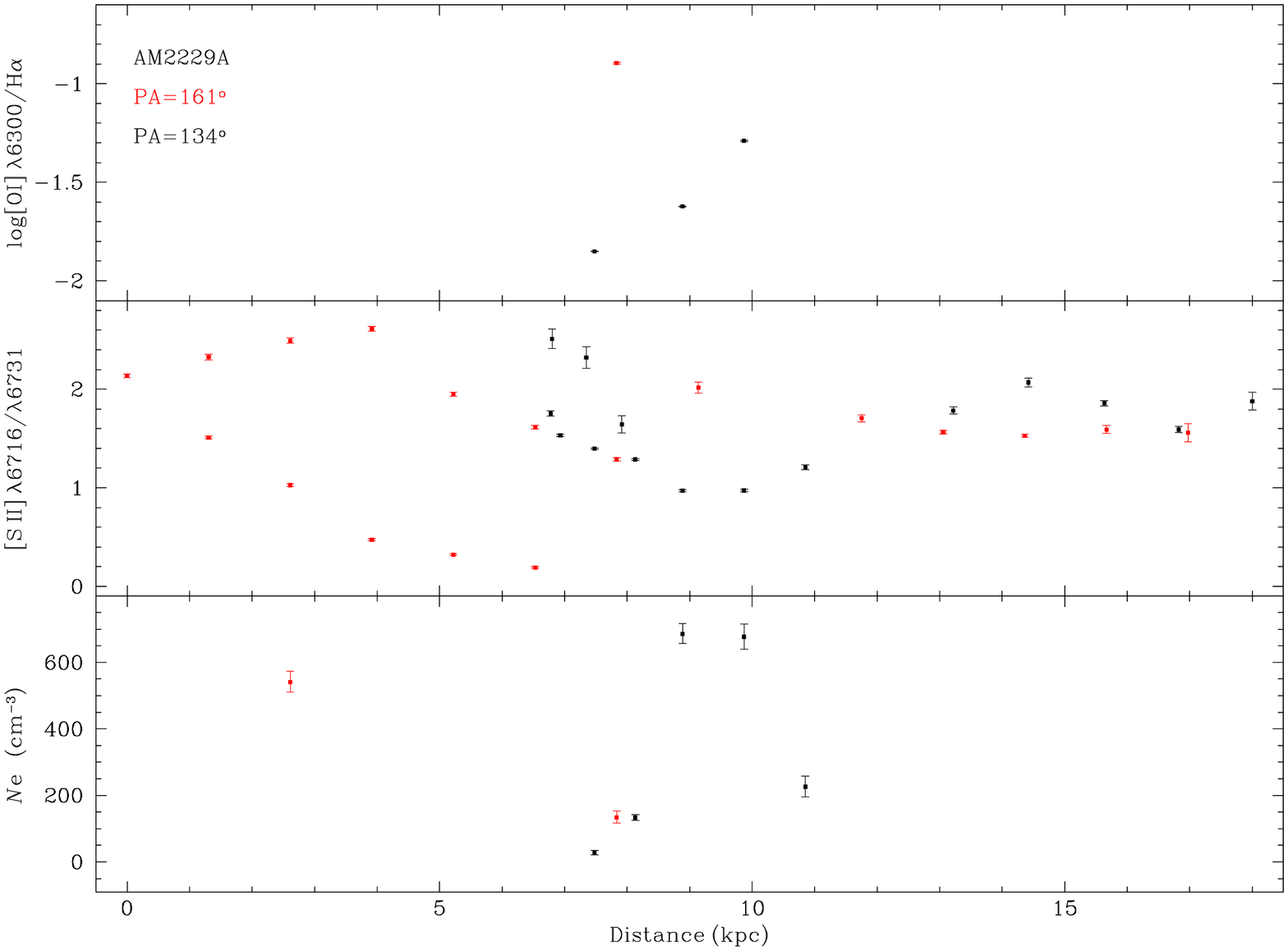}
\caption{Same as Fig. \ref{den_am1054}, but for AM\,2229A.}
\label{den_am2229a}
\end{figure*}

\subsection{AM\,2306-721}

AM\,2306-721 is a pair composed by a spiral galaxy  with disturbed arms (hereafter AM\,2306A) interacting with an irregular galaxy (hereafter AM\,2306B).
Both galaxies contain very luminous \ion{H}{ii} regions with H$\alpha$ luminosity in the range of  
8.30 $\times$10$^{39}$\,$\leq$ $L$\,(H$\alpha$) $\leq$ 1.32 $\times$10$^{42}$\,erg\,s$^{-1}$ and high star-formation rate in the
 range of 0.07 - 10  $M_\odot$ yr$^{-1}$,  as estimated from H$\alpha$  images by  \citet{ferreiro08}.

The few measurements of electron densities  provide values in the range of $N_{\rm e}= 32-298 $~$\mathrm{cm^{-3}}$
and $N_{\rm e}= 19-826 $~$\mathrm{cm^{-3}}$  for AM\,2302A and AM\,2306B,
respectively.  Although, we do not have estimates of the electron density at the centre of the main galaxy, the spatial profile seems to indicate  an increasing of the $N_{\rm  e}$ toward the centre of  the galaxy, which 
could be a consequence of gas inflow. 
Again, in the secondary galaxy, the electron density smoothly increases from
  about 4 kpc toward the outer regions of the galaxy  to
   the end of the spiral arm at the Southeast.

\begin{figure*}
\begin{center}
\includegraphics*[angle=-90,width=0.85\textwidth]{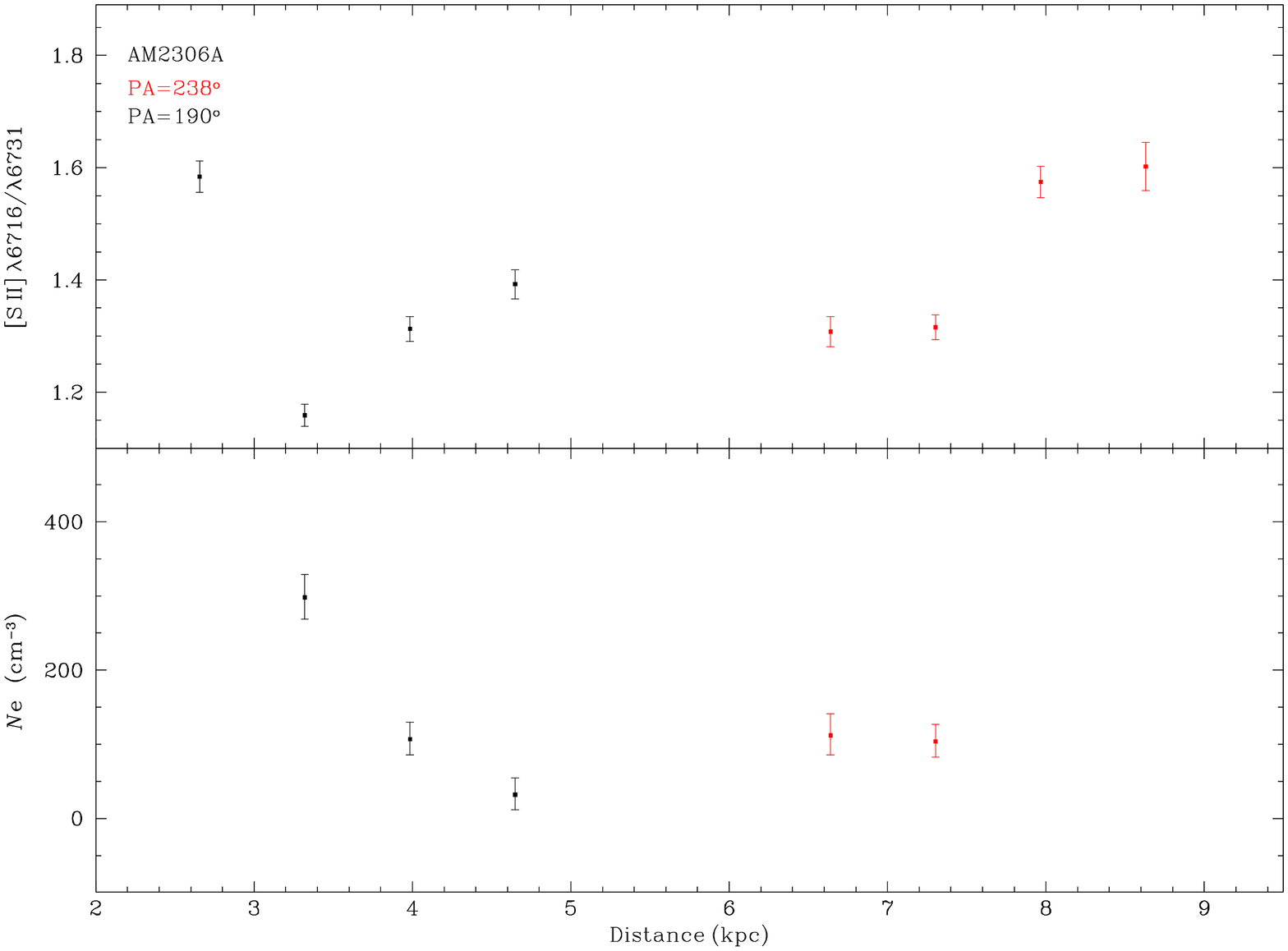}
\caption{Same as Fig. \ref{den_am1054}, but for AM\,2306A.}
\label{den_am2306a}
\includegraphics*[angle=-90,width=0.85\textwidth]{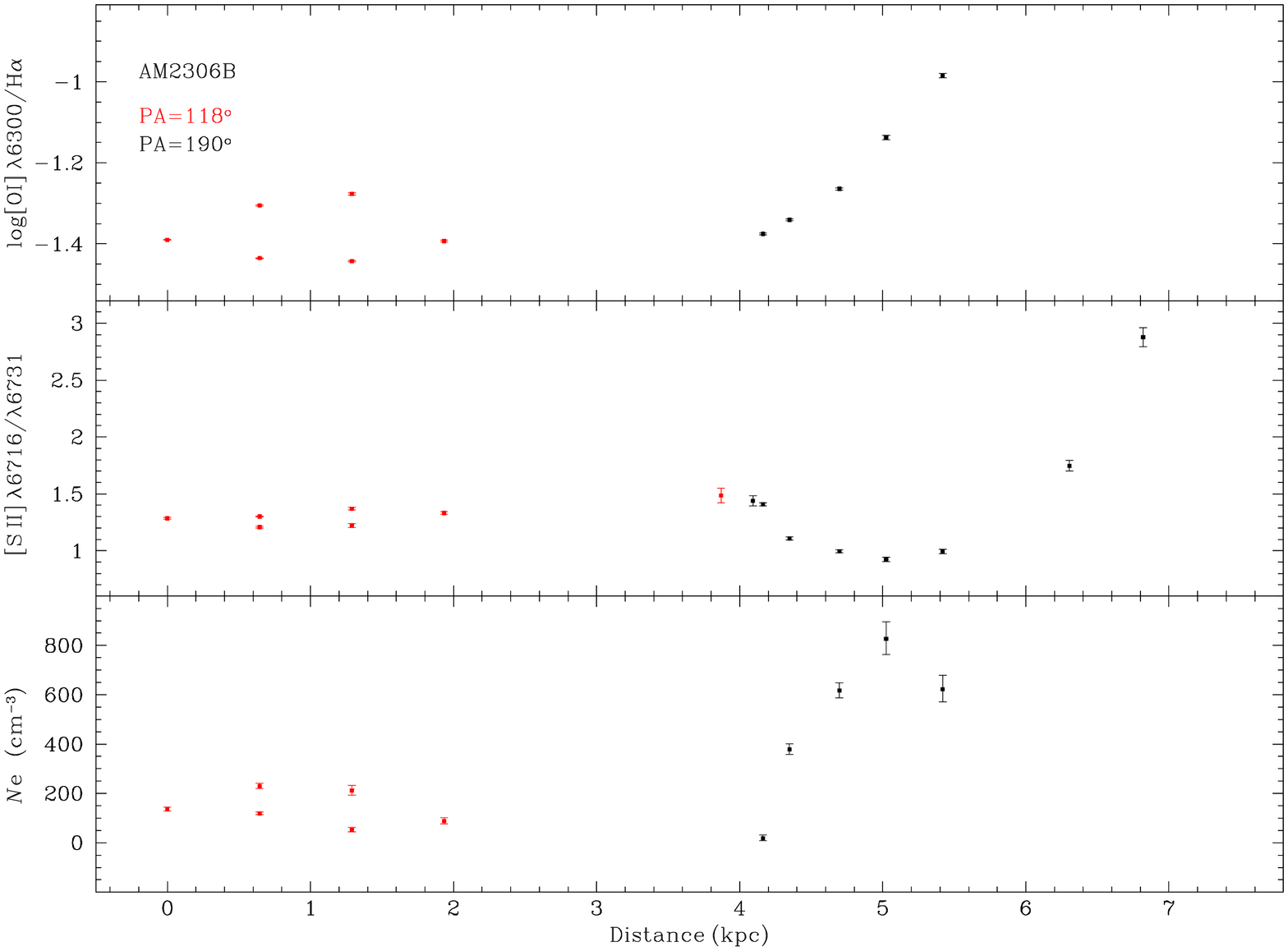}
\caption{Same as Fig. \ref{den_am1054}, but for AM\,2306B.}
\label{den_am2306b}
\end{center}
\end{figure*}

\subsection{AM\,2322-821}

AM\,2322-821 is composed of a SA(r)c galaxy with disturbed
arms (hereafter AM\,2322A) in interaction with an irregular
galaxy (hereafter AM\,2322B). Both galaxies contain very
luminous \ion{H}{ii} regions with
2.53$\times$10$^{39}$\,$\leq$ $L$\,(H$\alpha$) $\leq$ 1.45$\times$10$^{41}$\,erg\,s$^{-1}$  and
star formation rates from 0.02 to 1.15\,$M_{\odot}$\,yr$^{-1}$ \citep{ferreiro08}. 

The distribution of electron temperatures exhibits variations of very low
amplitude across the radius of AM\,2322A.  
One region (at about 2 kpc from the centre of the galaxy)  has  four
values of densities systematically higher than the other apertures along the  
radius of galaxy. This region is   marked in Fig.~\ref{am2322_zoom}. In this region, the
values of densities are in the range of $N_{\rm e}=803-1121$~$\mathrm{cm^{-3}}$.  We found 
a mean electron density of $N_{\rm e}= 200\pm12 $~$\mathrm{cm^{-3}}$.
AM\,2322B  presents a relatively homogeneous electron density distribution, with a mean density of 
$N_{\rm e}= 24\pm4.8 $~$\mathrm{cm^{-3}}$. This is the galaxy  with  the lowest density in our sample.

\begin{figure*}
\begin{center}
\includegraphics*[angle=-90,width=0.85\textwidth]{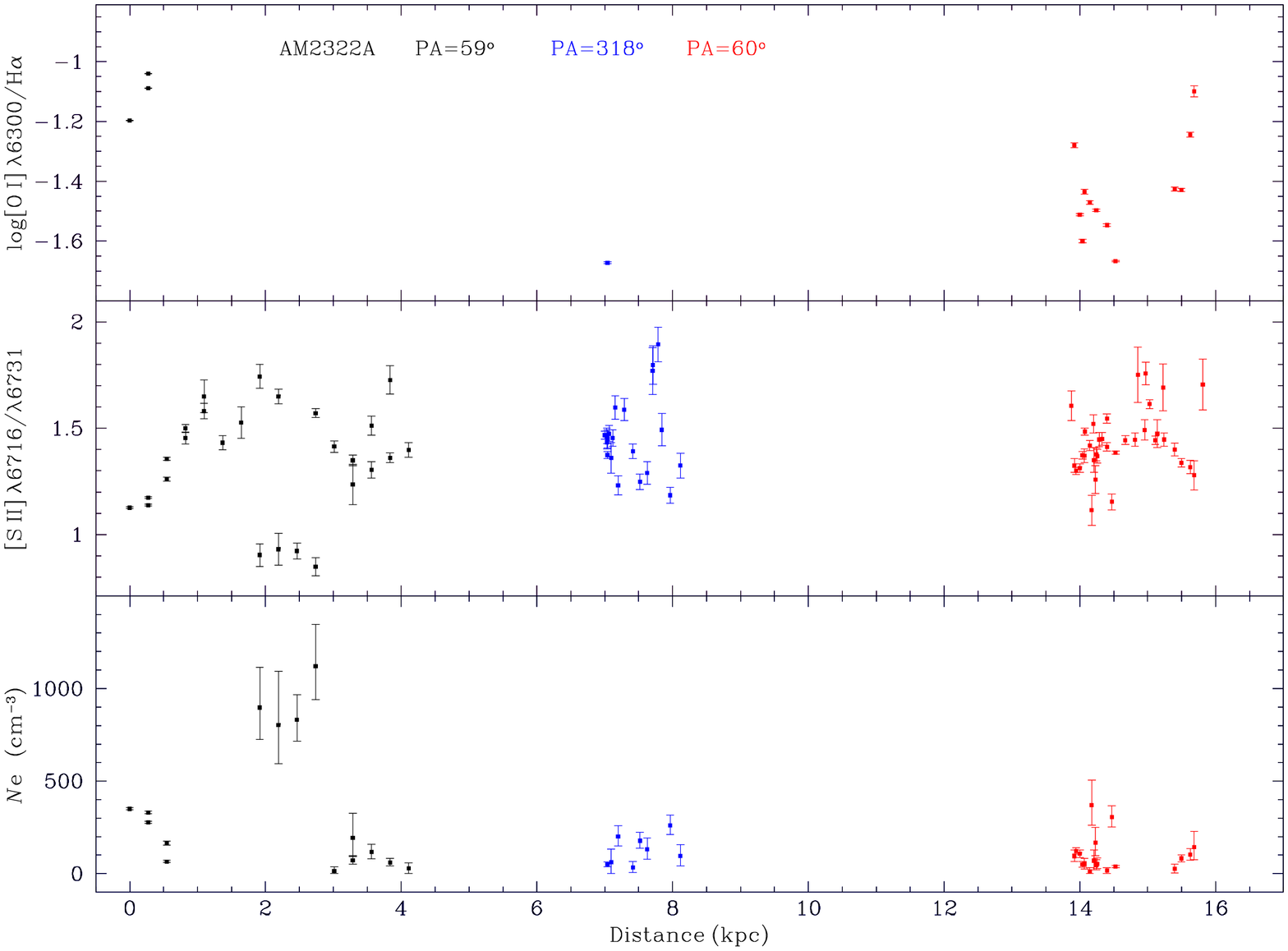}
\caption{Same as Fig. \ref{den_am1054}, but for AM\,2322A.}
\label{den_am2322a}
\includegraphics*[angle=-90,width=0.85\textwidth]{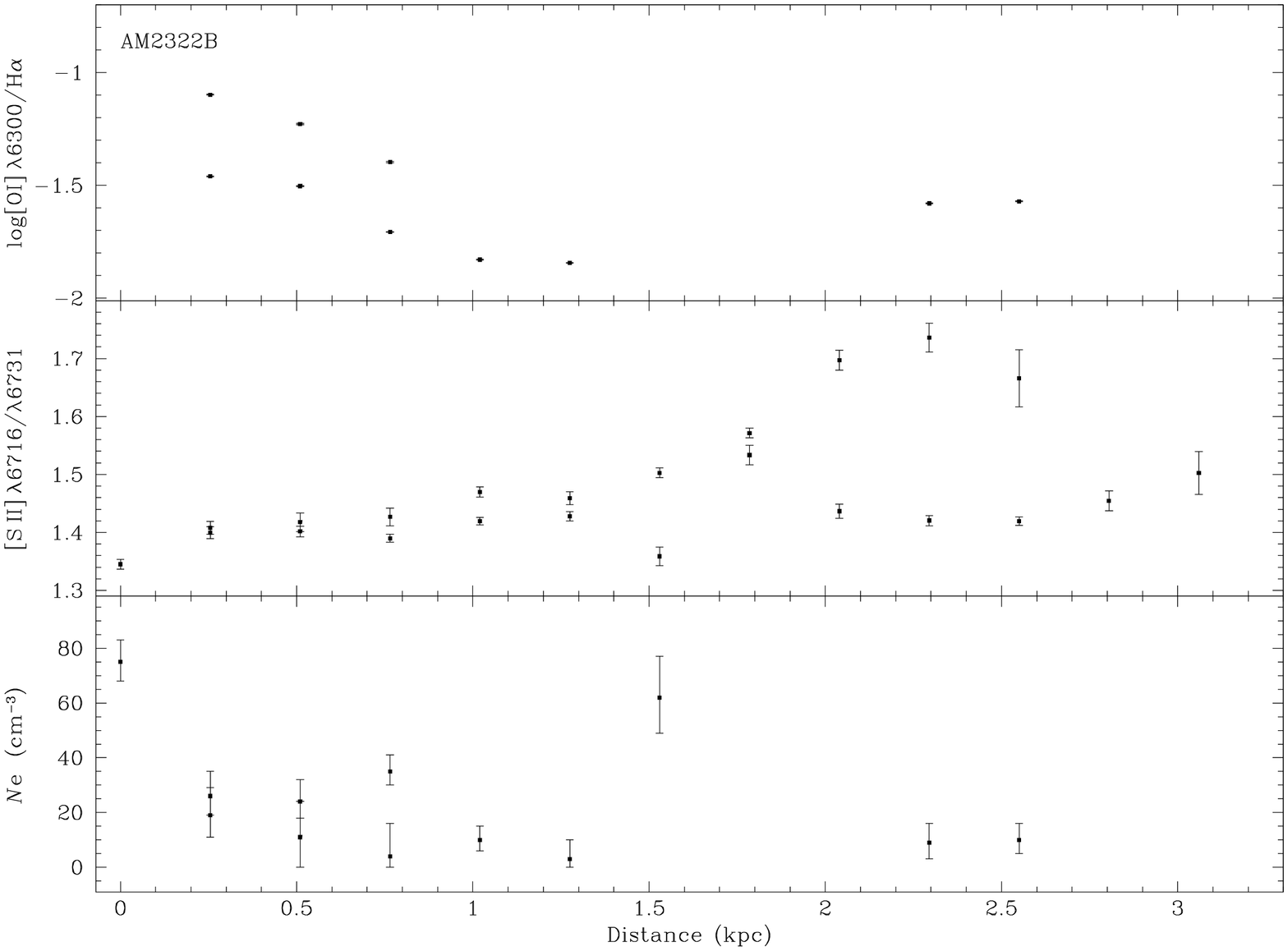}
\caption{Same as Fig. \ref{den_am1054}, but for AM\,2322B.}
\label{den_am2322b}
\end{center}
\end{figure*}

\begin{figure}
\begin{center}
\includegraphics*[width=\columnwidth]{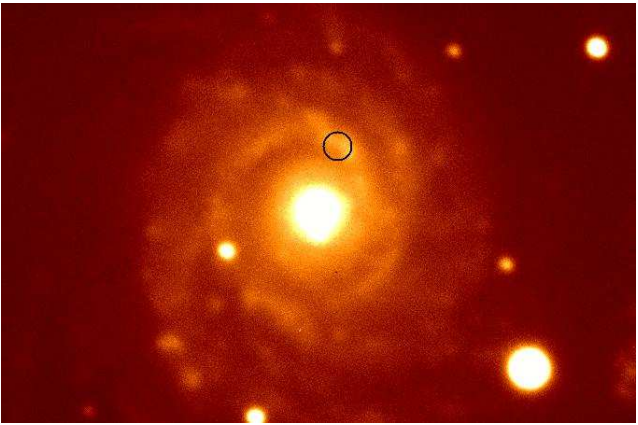}
\end{center}
\caption{Image of AM\,2322A  with the region of high density (see the text) marked with a circle.}
\label{am2322_zoom}
\end{figure}

\section{Discussion}
\label{discussion}

\begin{figure*}
\centering
\begin{center}
\includegraphics*[angle=-90, width=0.495\textwidth]{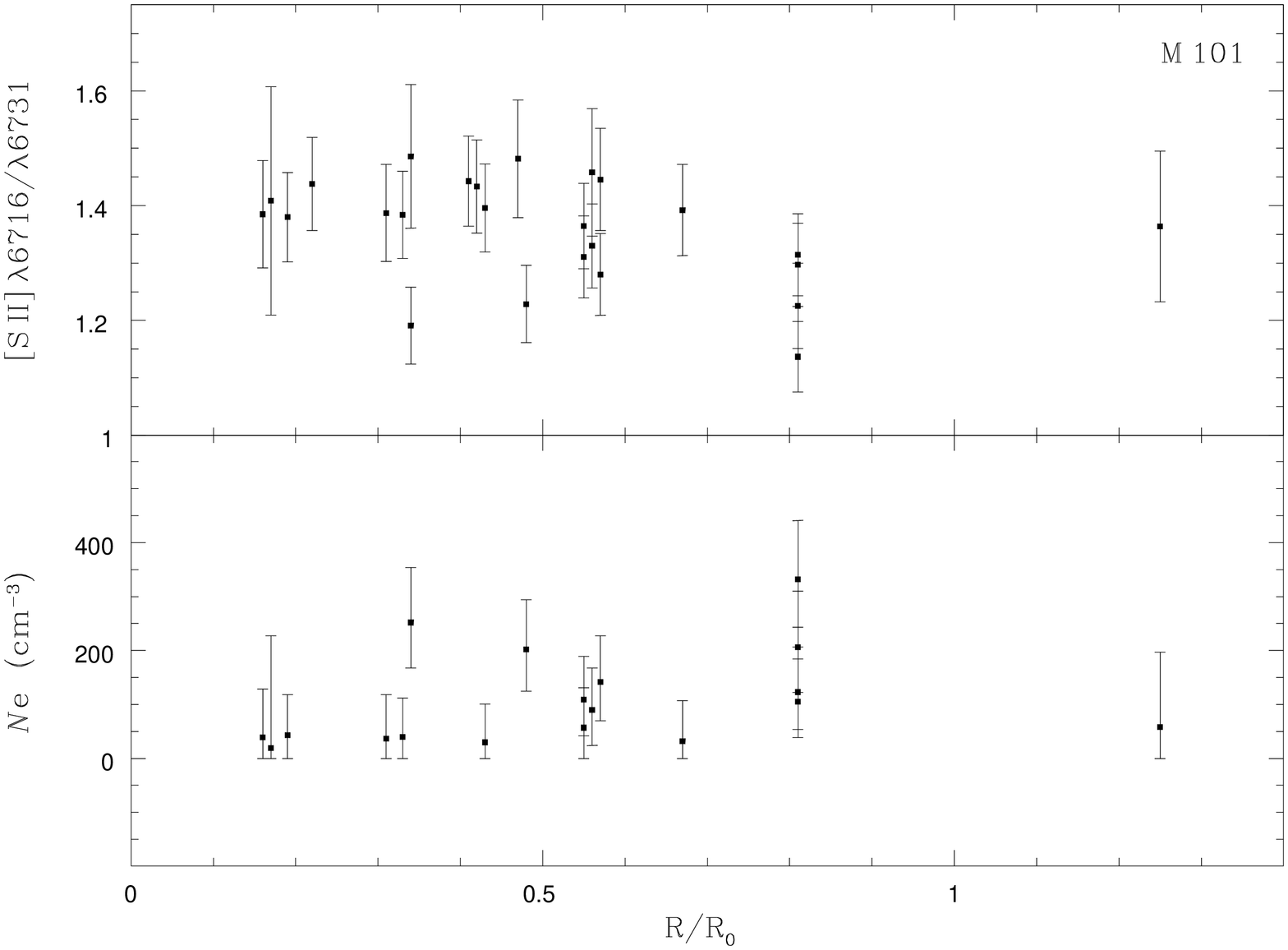}
\includegraphics*[angle=-90,width=0.495\textwidth]{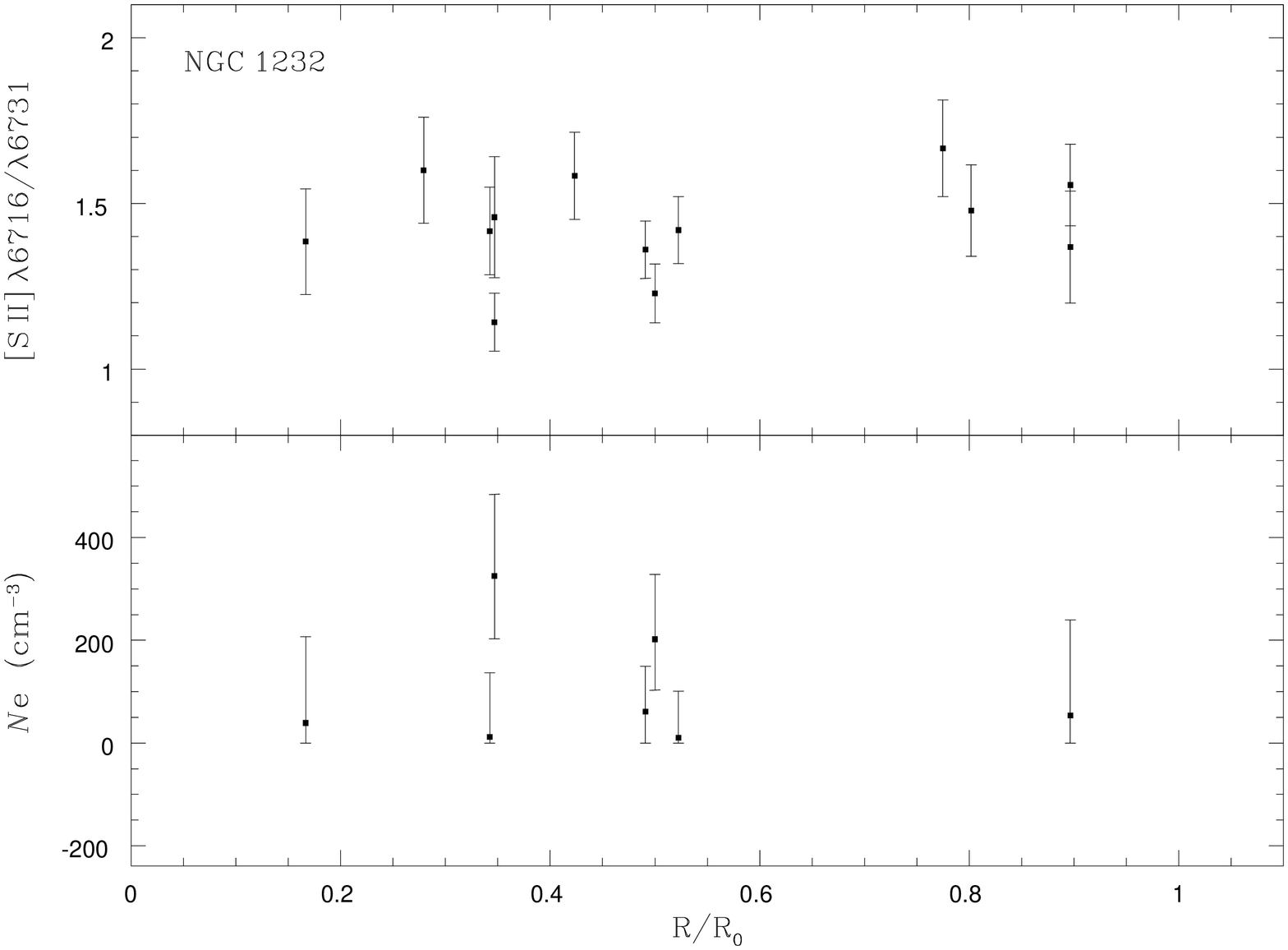}
\includegraphics*[angle=-90,width=0.495\textwidth]{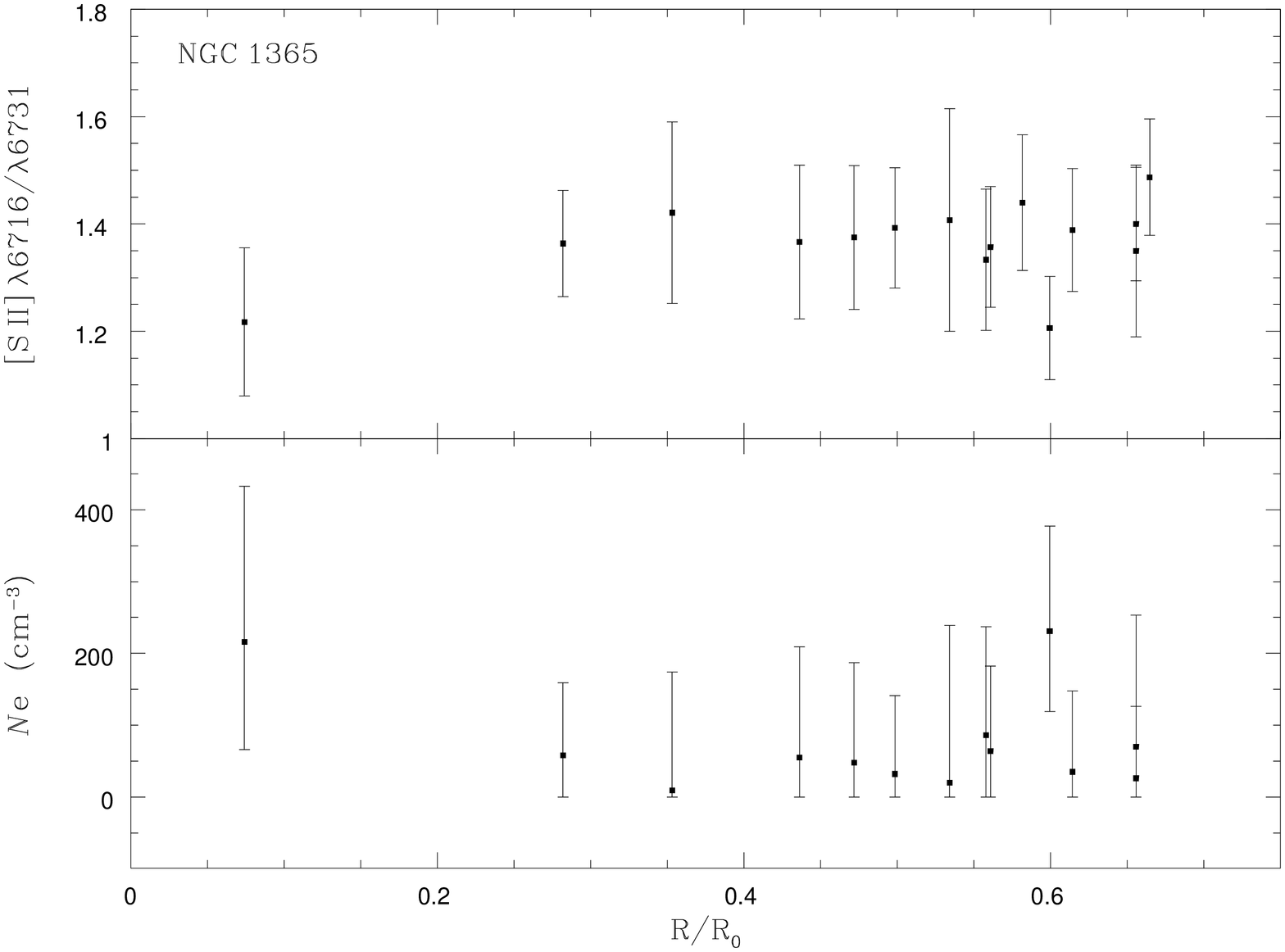}
\includegraphics*[angle=-90,width=0.495\textwidth]{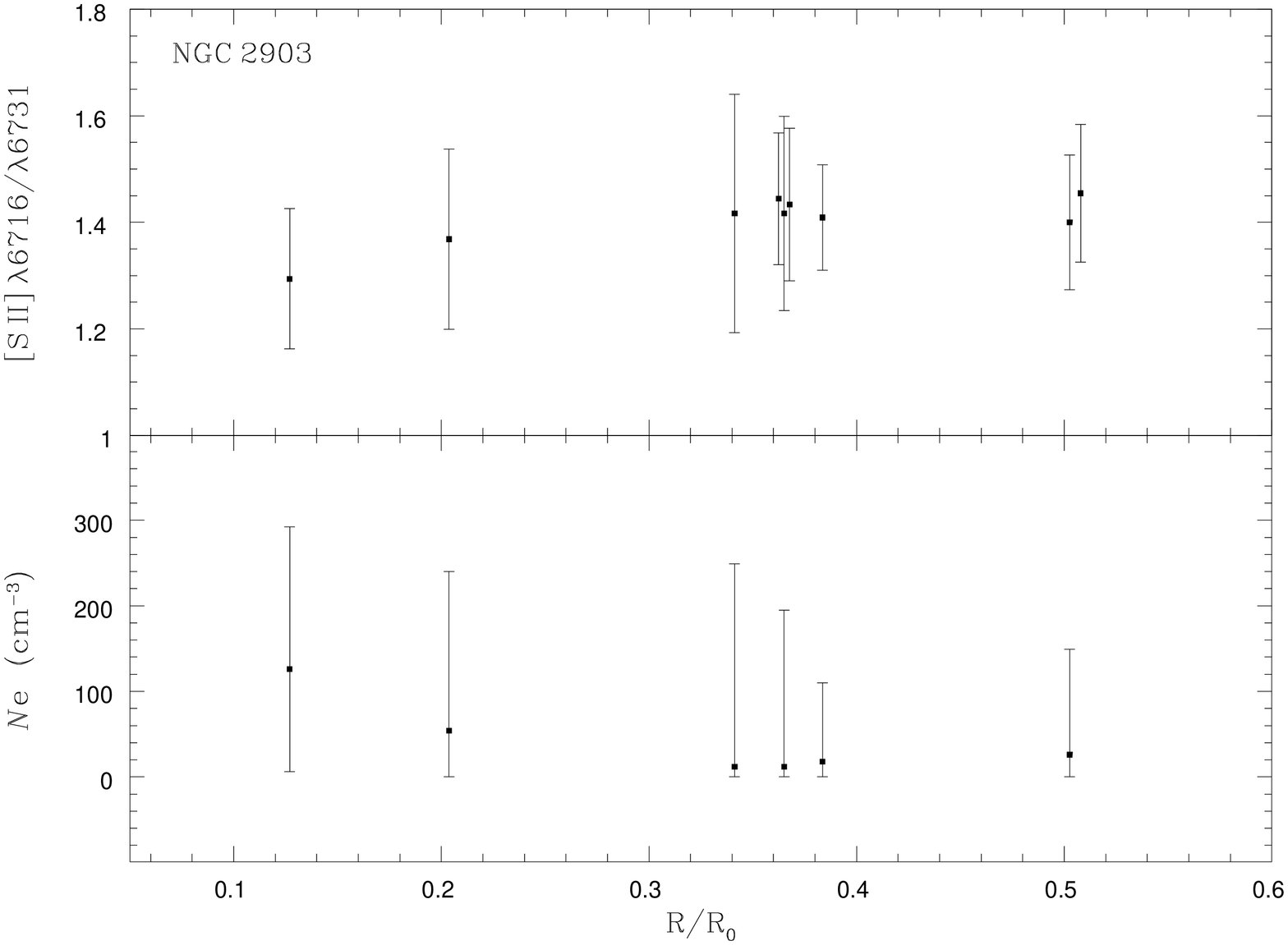}
\includegraphics*[angle=-90,width=0.495\textwidth]{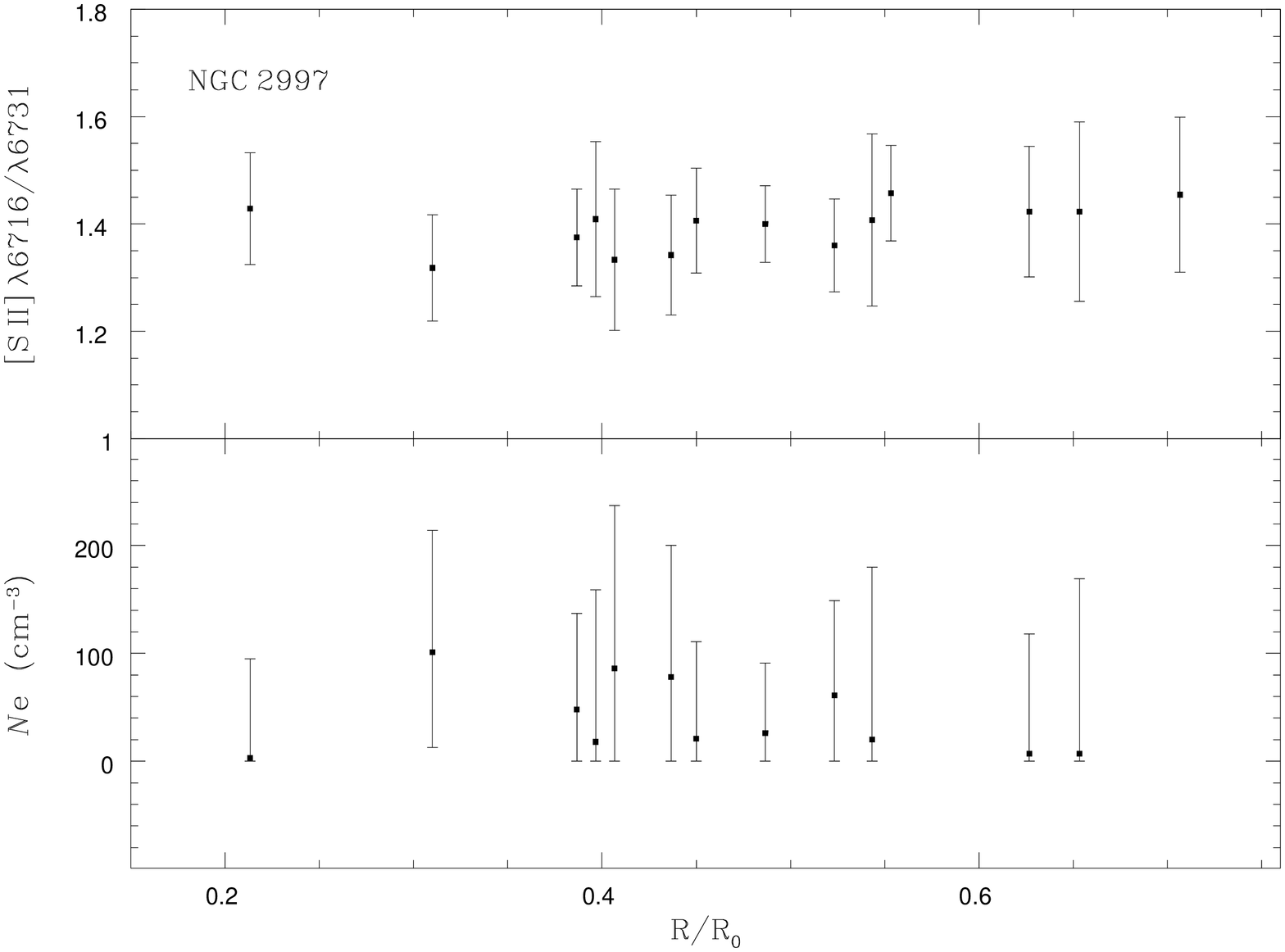}
\includegraphics*[angle=-90,width=0.495\textwidth]{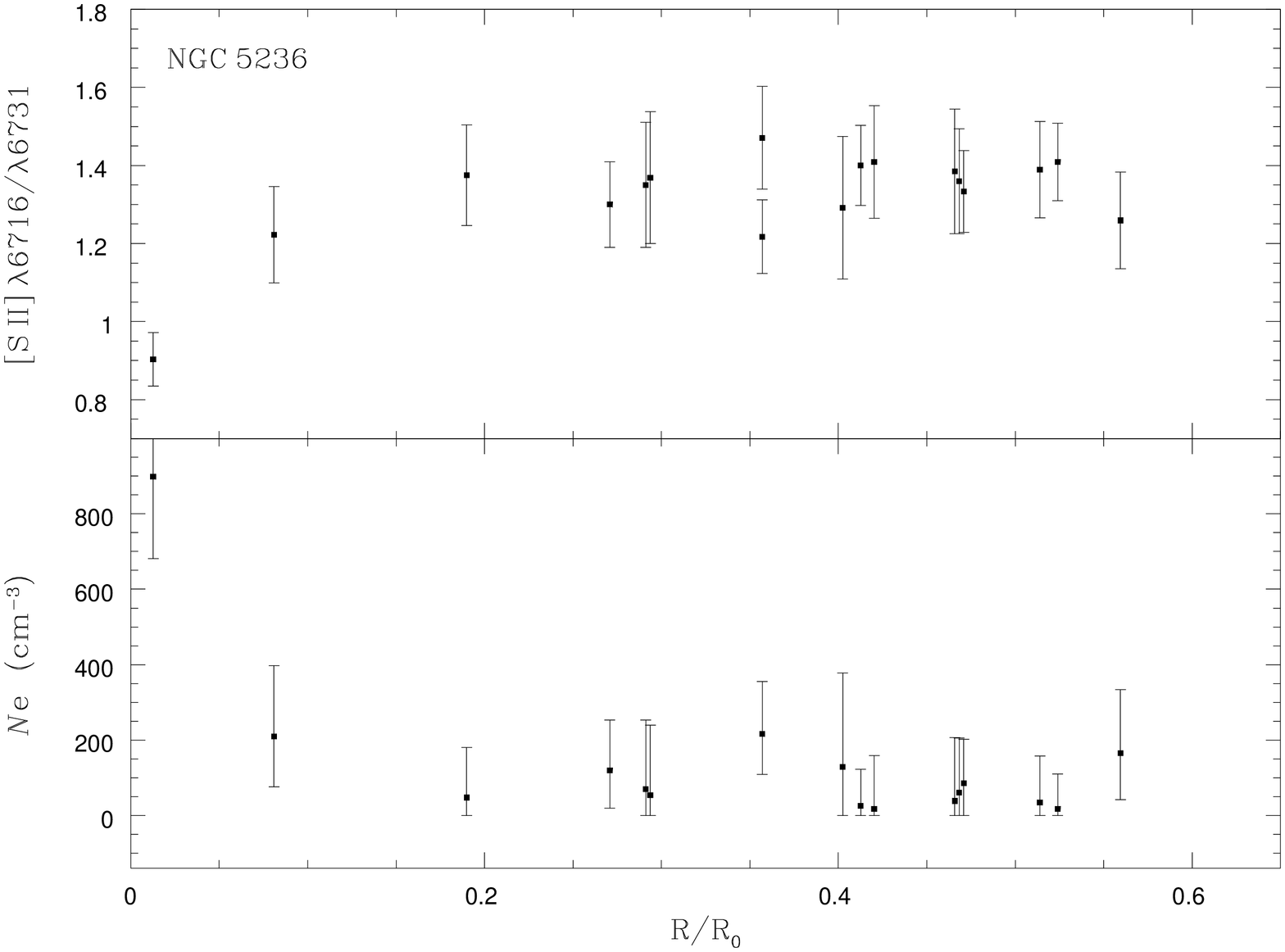}
\end{center}
\caption{Profiles of the electron density as a function of R/R$_{0}$,
  where R$_{0}$ is the galactocentric distance deprojeted for the isolated galaxies M\,101, NGC\,1232, NGC\,1365, NGC\,2903, NGC\,2997 and NGC\,5236.}
\label{m101}
\end{figure*}

To verify if there are  differences between the $N_{\rm e}$ values observed in
 the \ion{H}{ii} regions  of our sample  
and those obtained in isolated galaxies, we have calculated the electron densities from published  measurements of the
[\ion{S}{ii}] line-ratio  for disk \ion{H}{ii} regions  in the  isolated galaxies M\,101, NGC\,1232, NGC\,1365, NGC\,2903, NGC\,2997, and NGC\,5236
and compared these values with our results.
The data of these objects were taken from \citet{kennicutt03} for M\,101  and from  \citet{bresolin05} for the other galaxies.
The same atomic parameters and electron temperature  adopted for our  determinations were used.  
The spatial profiles of the  [S{\scriptsize\,II}]$\lambda$6716/$\lambda$6731 ratio and  the 
electron densities derived for  some \ion{H}{ii} regions in the isolated galaxies are shown in Fig.~\ref{m101}. 

As can be seen in  this figure, the estimated electron densities are relatively homogeneous along the radius of each isolated galaxy.
The  derived mean electron densities are in the range of 
$N_{\rm e}=40-137\, \rm cm^{-3}$.  Only one high value of 
$N_{\rm e}\approx900\:\rm cm^{-3}$ is derived in the central region of NGC\,5236.
It is a metal-rich  \ion{H}{ii} region, with a low electron temperature of $T_{\rm e}$(\ion{O}{iii})$=4\,000 \pm 2\,000$~K  and an oxygen abundance 
of 12+log(O/H)$\approx$\,8.9 dex as derived by   \citet{bresolin05}. This high
value can be  caused by  mass loss and strong stellar winds from 
embedded Wolf Rayet stars,  which are common in metal-rich environments (e.g.\ \citealt{pindao02, bresolin02, shaerer00}). 
If the adopted electron temperature is  $T_{\rm e}$(\ion{O}{iii})$=4\,000$ K, an estimation of $N_{\rm e}\approx623\:\rm cm^{-3}$ is obtained. This value is about 
30\% lower than the one obtained assuming an electron temperature of $T_{\rm e}$(\ion{O}{iii})$=10\:000$ K.  Then, even though the dependence 
of the $N_{\rm e}$ with the electron temperature is weak, it could have an important effect when temperature fluctuations of high amplitude were observed in 
\ion{H}{ii} regions.

The values of the electron density obtained from our sample of interacting
galaxies are systematically higher than those derived for the isolated ones. The mean electron density values  derived by us for the
  interacting galaxies in our sample
are in the range of $N_{\rm e}=24-532\, \rm cm^{-3}$,  which also show higher
values than for isolated galaxies. \citet{newman12}, for the
clumpy star-forming galaxy  ZC406690, also obtained high electron density
values ($N_{\rm e}=300-1800\, \rm cm^{-3}$).
Moreover, several  of our
interacting galaxies  (AM\,2306B, AM\,1219A, and AM\,1256B)
show a slight increment of the $N_{\rm e}$ in the outer parts of the galaxy, opposite of  what is observed in the isolated galaxies, where the electron density
is homogeneous along the radius.  The high electron density  values found in the outlying parts  for the majority of the
objects of our sample would be due to zones of induced star formation by direct cloud-cloud interaction
(for a review see \citealt{bournaud11}). In these  regions,   turbulent flows can locally compress the
gas,   forming over-densities  that subsequently cool and collapse into star-forming clouds \citep{duc13, elmegreen02}. 
Although we do not have estimates of the electron density  at  the centre of AM\,2306A,  the spatial profile 
seems to indicate an increasing of $N_{\rm  e}$ toward the centre of  the galaxy, which  could be due to inflowing gas. However, 
in only a  few regions in this galaxy were possible to estimate $N_{\rm e}$, therefore, this is a marginal conclusion.

 It is worth mentioning that \ion{H}{ii} regions seemed to be inhomogeneous, and the zones where most of the emission from the ionized gas is originated only occupy a small fraction of the total volume (i.e., small filling factor). Hence, our electron density
values derived from the [\ion{S}{ii}] emission lines are representative of a fraction of the total volume of the
\ion{H}{ii} region (referred as \textit{in situ} electron densities).
According to \citet{giammanco04}, these inhomogeneities, if optically thick, 
can modify the  determinations of electron temperatures and densities, ionization parameters, and  abundances.  \citet{copetti00} presented a study on  internal variation of the electron density
 in a sample of spatially resolved  galactic \ion{H}{ii} regions of different sizes and evolutionary stages.  These authors found
 that the electron density within  \ion{H}{ii} regions (e.g. S\,307) can range from about 30 to 600 $\rm cm^{-3}$, and 
 a filling factor of the order of  0.1  is compatible with their data.  Therefore, the estimated electron densities could be about 10\% of the in situ values sampled by the sulphur line ratio.

 \begin{figure}
\centering
\begin{center}
\includegraphics*[angle=-90,width=\columnwidth]{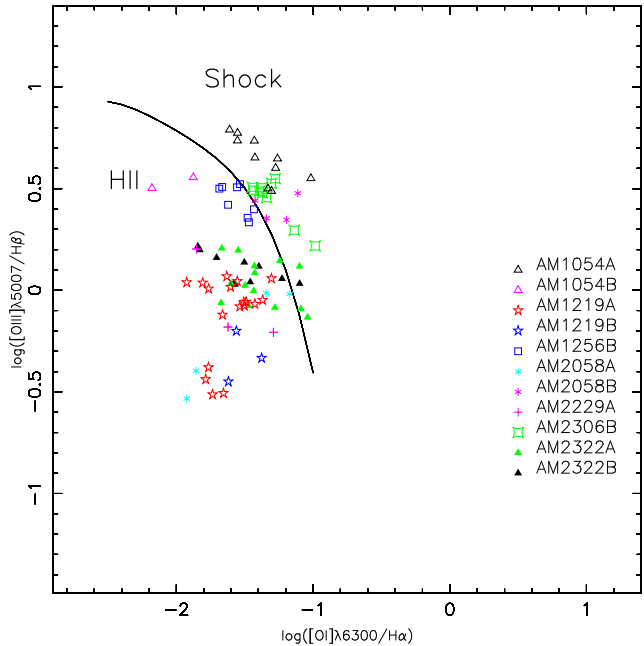}
\end{center}
\caption{Diagnostic diagram of [O{\scriptsize\,III}]$\lambda$5007/H$\beta$ vs. [O{\scriptsize\,I}]$\lambda$6300/H$\alpha$.  The black solid line from  \citet{kewley06} separates the objects ionized by massive stars from the ones containing active nuclei and/or shock
excited gas.   The data for  distinct  galaxies are marked by different symbols as indicated.
The typical error bar (not shown) of the emission line ratios is about 10 per cent.}
\label{choque}
\end{figure}

An important  issue is  to study the origin of the high electron density
values found in the \ion{H}{ii} regions of our sample. The  presence of gas shock excitation
in interacting galaxies 
is very important not only because they affect quantities derived
from spectroscopy, but also  due to they act as a  mechanism for
dissipating the kinetic energy and  the angular momentum of the
infalling gas in merging systems, as discussed  by \citet{rich11}. 
 The gas shock also increases the density due to
the compression of the interstellar material.
To  analyse if  the presence of shock-excited gas produces the high electron
density values,  
the diagnostic diagram [O{\scriptsize\,III}]$\lambda$5007/H$\beta$ vs.\
[O{\scriptsize\,I}]$\lambda$6300/H$\alpha$ proposed by \citet{baldwin81} and \citet{veilleux87},  and used to separate objects ionized by stars,  
by shocks and/or  active nuclei 
(AGN) was considered.

In Fig.~\ref{choque},  the diagnostic diagram containing the data of the  \ion{H}{ii} regions  studied by us is shown. 
The galaxy nuclei data   are not shown in this diagram.
We also show in this plot, the line proposed by \citet{kewley06} to
separate objects with distinct ionizing sources: shock gas and
massive star excitations.  
We can see that  the all \ion{H}{ii} regions in  AM\,1054A,
AM\,2058B, AM\,2306B, and  some regions  in AM\,2322A (3 apertures) and
AM\,2322B (1 aperture) occupy  the area where objects with shock as the
main ionizing source are located.  
 The number of objects represented in Fig.~\ref{choque} differ from those in the profile figures (Figs.~\ref{den_am1054}-\ref{den_am2322b}) because the 
[\ion{O}{iii}]$\lambda5007$/H$\beta$ ratio could not be measured for all
apertures.
From the comparison of the spatial profiles of the electron density and
  the logarithm of [\ion{O}{i}]$\lambda6300$/H$\alpha$ in the AM\,1054A, 
  AM\,2058B, and AM\,2306B galaxies (Figs.~\ref{den_am1054}, \ref{den_am2058b}
  and \ref{den_am2306b}, respectively) we can note the following: 

\begin{itemize}
 \item AM\,1054A:  All regions of this galaxy have gas shock excitation and the values of electron
density are relatively high.

\item AM\,2058B: It is a  small galaxy and only a few apertures could be
  extracted.  As can be seen in  Fig.~\ref{choque},   all four disk \ion{H}{ii} regions
  of this galaxy have gas shock excitation, and from  Fig.~\ref{den_am2058b}, 
 we can note that these regions present low electron density values ($< 200\: \rm cm^{-3}$).

\item AM\,2306B: the regions with highest
  [O{\scriptsize\,I}]$\lambda$6300/H$\alpha$ and $N_{\rm e}$ values ($\approx 700$ $\rm cm^{-3}$ ) lie in the outskirts of galaxy. 
As can be seen in Fig.~\ref{den_am2306b}, it seems to be a trend in this
object: from about 4 kpc both the $N_{\rm e}$ and the [O\,I]/H$\alpha$ ratio
increase to the outer parts of the galaxy;  in the inner part (up to 2
  kpc), the profiles of these two quantities are almost flat showing low
  values. 

\end{itemize}

\begin{figure}
\centering
\includegraphics[angle=-90,width=1\columnwidth]{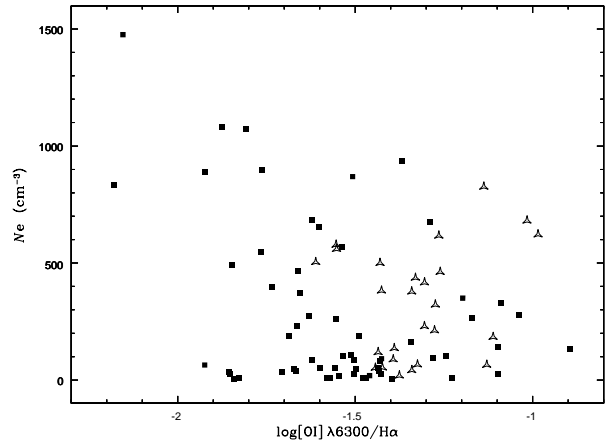}
\caption{Electron density values $N_{\rm e}$ 
derived  for our sample versus the observed
[\ion{O}{i}]$\lambda$6300/H$\alpha$  ratio.
Squares represent regions ionized by massive stars while triangles represent
those with 
gas shock excitation, according to the diagnostic diagram presented in Fig.~\ref{choque}.}
\label{fig33}
\end{figure}

 The cause of the high electron density values associated with the shock excitation region
 in interacting galaxies   is essential to understand  how the flux gas works in them. 
High-velocity  gas motions can destroy molecular clouds and quench star
formation \citep{tubbs82}. To investigate if the high electron density values found in our sample are
associated with the  presence of excitation by gas shock,
we plotted  in Fig.~\ref{fig33} the $N_{\rm e}$ versus the logarithm of
  the observed [\ion{O}{i}]$\lambda$6300/H$\alpha$ emission line
  ratio.  Objects with distinct gas excitation source, in according to Fig.~\ref{choque}, are indicated by different symbols.
No correlation is obtained   between the presence of shocks and electron densities. The highest electron density values found in our sample do not belong to objects with gas shock excitation. Therefore, 
the high electron density values found in the \ion{H}{ii} regions of our sample   do not seem to be caused 
by the  presence of gas shock excitation.  However, a deeper analysis such as investigating the presence of correlation between the velocity dispersion of some emission line and its intensity (e.g. \citealt{thaisa07}) or the implications of multiple kinematical components in the emission line profiles on the derived properties \citep{hagele13,hagele12,amorin12} is necessary to confirm our result.
Interestingly, the objects with the highest electron density values present the smallest
  [\ion{O}{i}]$\lambda$6300/H$\alpha$   line intensity ratios.

\section{Conclusions}
\label{conc} 
 An observational study of the effects of the interaction on the electron densities from the  \ion{H}{ii} regions along the radius of a sample of  interacting galaxies is performed. The data   consist of long-slit spectra of high signal-to-noise ratio in the 4390-7250 \AA{} obtained with the Gemini Multi-Object Spectrograph at Gemini South (GMOS).
The electron density was determined using the ratio of lines [S{\scriptsize\,II}]$\lambda$6716/$\lambda$6731.  The main findings are the following:

\begin{itemize}

\item
The electron density estimates obtained for some \ion{H}{ii} regions of our
sample of interacting galaxies are systematically higher than those derived for isolated
galaxies in the literature. The mean electron density values of interacting galaxies are in the range of $N_{\rm e}=24-532\, \rm cm^{-3}$, while those obtained for isolated galaxies are in the range of $N_{\rm e}=40-137\, \rm cm^{-3}$.

\item
Some interacting galaxies: AM\,2306B, AM\,1219A, and AM\,1256B show an increment of 
$N_{\rm  e}$ toward the outskirts of each system. This kind of relation is not observed in
isolated galaxies, where the electron density profile is rather flat along the radius of each
galaxy. 

\item
 The galaxies where the mechanism of gas shock excitation is present in almost all 
the \ion{H}{ii} regions are AM\,1054A, AM\,2058B, and  AM\,2306B. For the remaining galaxies, only  few
\ion{H}{ii} regions has emission lines excited by shocks, such as in AM\,2322B (1 point) and AM\,2322A (4 points).
It is noteworthy that only in three of all objects analysed here the main
  excitation mechanism for all of their \ion{H}{ii} regions is shocks. 
  
\item
 No correlation is obtained  between the presence of shocks and electron densities. Indeed,  the highest electron density values found in our sample do not belong to the objects with  gas shock excitation. Therefore, 
the high electron density values found in the \ion{H}{ii} regions of our sample do not seem to be caused by the presence of gas shock excitation.

\end{itemize}

\section*{Acknowledgements}

Based on observations obtained at the Gemini Observatory, which is operated by 
the Association of Universities for Research in Astronomy, Inc., under a 
cooperative agreement with the NSF on behalf of the Gemini partnership: the 
National Science Foundation (United States), the Science and Technology
Facilities Council (United Kingdom), the National Research Council (Canada), CONICYT
(Chile), the Australian Research Council (Australia), 
Minist\'erio da Ciencia e Tecnologia (Brazil), and SECYT (Argentina).

A. C. Krabbe, O. L. Dors Jr., and D. A. Rosa  thank the support of FAPESP, process 2010/01490-3,   2009/14787-7, and 2011/08202-6 respectively.

We also thank Ms. Alene Alder-Rangel for editing the English in this manuscript.


\begin{thebibliography}{99}

\bibitem[\protect\citeauthoryear{Allen et al.}{2008}]{allen08} Allen, M., Brent, A., Dopita, M. A. 2008, ApJS, 178, 20

\bibitem[\protect\citeauthoryear{Amor{\'{\i}}n et 
al.}{2012}]{amorin12} Amor{\'{\i}}n R., V{\'{\i}}lchez J.~M., 
H{\"a}gele G.~F., Firpo V., P{\'e}rez-Montero E., Papaderos P., 2012, ApJ, 
754, L22 


\bibitem[\protect\citeauthoryear{Baldwin et al.}{1981}]{baldwin81} Baldwin, J.~A., Phillips, M.~M., Terlevich, R. 1981, PASP, 93, 5

\bibitem[\protect\citeauthoryear{Bournaud}{2011}]{bournaud11} Bournaud, F. 2011, EAS Publications Series, 51,
107


\bibitem[\protect\citeauthoryear{Bowen et al.}{1960}]{bowen60} Bowen, I. S., 1960, ApJ, 132, 1.

\bibitem[\protect\citeauthoryear{Bresolin et al.}{2005}]{bresolin05} Bresolin, F., Schaerer, D.,  Gonz\'alez Delgado, R.~M.,  Stasi\'nska, G. 2005, A\&A, 441, 981

\bibitem[\protect\citeauthoryear{Bresolin \& Kennicutt}{2002}]{bresolin02} Bresolin, F., \& Kennicutt, R.~C. 2002, ApJ, 572, 838

\bibitem[\protect\citeauthoryear{Casta\~neda et al.}{1992}]{castaneda92} Casta\~neda, H.~O., V\'{\i}lchez,  J.~M.,  Copetti, M.~V.~F. 1992, A\&A, 260, 370

\bibitem[\protect\citeauthoryear{Copetti et al.}{2000}]{copetti00}  Copetti, M.~V.~F.,  Mallmann, J.~A.~H., Schmidt, A.~A.; Casta\~neda, H.~O. 2000, A\&A, 357, 621

\bibitem[\protect\citeauthoryear{Copetti \& Writzl}{2002}]{copetti02}  Copetti, M.~V.~F.,  \& Writzl, B.~C.  2002, A\&A, 382, 282

\bibitem[\protect\citeauthoryear{de Vaucouleurs et al.}{1991}]{devaucouleurs91} de Vaucouleurs, G., de Vaucouleurs, A., Corwin, H.~G., Jr., et al. 1991, Book-Review - Third Reference Catalogue of Bright Galaxies Sky and Telescope, 82, 621 

\bibitem[\protect\citeauthoryear{Donzelli  \& Pastoriza}{2000}]{donzelli00} Donzelli, C.~J., Pastoriza, M.~G. 2000, AJ, 120, 189

\bibitem[\protect\citeauthoryear{Donzelli  \& Pastoriza}{1997}]{donzelli97} Donzelli, C.~J., Pastoriza, M.~G. 1997,  ApJS, 111, 181

%\bibitem[\protect\citeauthoryear{Dopita et al.}{2002}]{dopita02} Dopita, M.~A., Pereira, M., Kewley, L.~J., Capaccioli, M. 2002, ApJ, 47, 72

\bibitem[\protect\citeauthoryear{Dopita et al.}{2000}]{dopita00} Dopita, M.~A.,  Kewley, L.~J., Heisler, C.~A., Sutherland, R.~S. 2000, ApJ., 542, 224

\bibitem[\protect\citeauthoryear{Dors et al.}{2011}]{dors11} Dors J. O. L., Krabbe A., H¨agele G. F., P´erez-Montero E., 2011, MNRAS, 415, 3616

\bibitem[\protect\citeauthoryear{Dors et al.}{2008}]{dors08}  Dors, O.~L., Storchi-Bergmann, T., Riffel, R.~A., Schimdt, A.~A. 2008, A\&A, 482, 59

\bibitem[\protect\citeauthoryear{Duc et al.}{2013}]{duc13} Duc, P.-A., Belles, P.-E.,  Brinks, E.,  Bournaud, F. 2013, Proceedings of the International Astronomical Union,
292, 323

\bibitem[\protect\citeauthoryear{Elmegreen}{2002}]{elmegreen02} Elmegreen, B.~G. 2002,  ApJ,  577, 206

\bibitem[\protect\citeauthoryear{Ferreiro  \& Pastoriza}{2004}]{ferreiro04} Ferreiro, D.~L.,  Pastoriza, M.~G. 2004, A\&A, 428, 837

\bibitem[\protect\citeauthoryear{Ferreiro et al.}{2008}]{ferreiro08} Ferreiro, D.~L.,  Pastoriza, M.~G., Rickes, M. 2008, A\&A, 481, 645


\bibitem[\protect\citeauthoryear{Giammanco et al.}{2004}]{giammanco04} Giammanco, C., Beckman, J.~E., Zurita, A., \& Rela{\~n}o, M.\ 2004, A\&A, 424, 877 
\bibitem[\protect\citeauthoryear{H{\"a}gele et 
al.}{2012}]{hagele12} H{\"a}gele G.~F., Firpo V., Bosch G., 
D{\'{\i}}az {\'A}.~I., Morrell N., 2012, MNRAS, 422, 3475 

\bibitem[\protect\citeauthoryear{H{\"a}gele et al.}{2013}]{hagele13} H{\"a}gele G.~F., D{\'{\i}}az {\'A}.~I., 
Terlevich R., Terlevich E., Bosch G.~L., Cardaci M.~V., 2013, MNRAS, 432, 
810 

\bibitem[\protect\citeauthoryear{Huchra et al.}{2012}]{huchra12} Huchra, J.~P., Macri,  L.~M., Masters, K.~L., et al.\ 2012, ApJS, 199, 26 

\bibitem[\protect\citeauthoryear{Jones et al.}{2009}]{jone09} Jones, D.~H., Read,  M.~A., Saunders, W., et al.\ 2009, MNRAS, 399, 683 

\bibitem[\protect\citeauthoryear{Keenan et al.}{1993}]{keenan93} Keenan, F. P. and Hibbert, A. and Ojha, P. C. and Collon, E. S. 1993, Phys. Scr. A., 47, 48-129.

\bibitem[\protect\citeauthoryear{Kennicutt et al.}{2003}]{kennicutt03} Kennicutt, R.~C.,  Bresolin, F.,  Garnett, D.~R. 2003, ApJ, 591, 801

\bibitem[\protect\citeauthoryear{Kennicutt, Keel, 
\& Blaha}{1989}]{kennicutt89} Kennicutt R.~C., Jr., Keel W.~C., Blaha C.~A., 1989, AJ, 97, 1022

\bibitem[\protect\citeauthoryear{Kennicutt}{1984}]{kennicutt84} Kennicutt, R.~C. 1984, ApJ, 287, 116

\bibitem[\protect\citeauthoryear{Kewley et al.}{2006}]{kewley06} Kewley, L.~J., Groves,  B., Kauffmann, G., \& Heckman, T.\ 2006, MNRAS, 372, 961 

\bibitem[\protect\citeauthoryear{Kewley et al.}{2010}]{kewley10}  Kewley L. J., Rupke D., Jabran Hahid H., Geller M. J., Barton E. J., 2010, ApJ, 721, L48

\bibitem[\protect\citeauthoryear{Kewley \& Dopita}{2002}]{kewley02}  Kewley L. J., \& Dopita, M.~A. 2002, ApJSS, 145, 35

\bibitem[\protect\citeauthoryear{Krabbe et al.}{2011}]{krabbe11} Krabbe, A.~C., Pastoriza, M.~G.,  Winge, Cl\'audia,  Rodrigues, I.,  Dors, O. L.,  Ferreiro, D. L. 2011, MNRAS, 416, 38

\bibitem[\protect\citeauthoryear{Krabbe et al.}{2008}]{krabbe08} Krabbe, A. C.; Pastoriza, M. G.; Winge, Cl�udia; Rodrigues, I.; Ferreiro, D. L.

\bibitem[\protect\citeauthoryear{Lauberts \& Valentijn}{1989}]{lauberts89} Lauberts, A., \& Valentijn, E.~A.\ 1989, The Messenger, 56, 31 

\bibitem[\protect\citeauthoryear{Lauberts}{1982)}]{lauberts82} Lauberts, A.\ 1982, Garching:  European Southern Observatory (ESO), 1982,  

\bibitem[\protect\citeauthoryear{Lagos et al.}{2009}]{lagos09} 
Lagos P., Telles E., Mu{\~n}oz-Tu{\~n}{\'o}n C., Carrasco E.~R., Cuisinier 
F., Tenorio-Tagle G., 2009, AJ, 137, 5068

\bibitem[\protect\citeauthoryear{L{\'o}pez-Hern{\'a}ndez et 
al.}{2013}]{lopez13} L{\'o}pez-Hern{\'a}ndez J., Terlevich E., 
Terlevich R., Rosa-Gonz{\'a}lez D., D{\'{\i}}az {\'A}., 
Garc{\'{\i}}a-Benito R., V{\'{\i}}lchez J., H{\"a}gele G., 2013, MNRAS, 
430, 472

\bibitem[\protect\citeauthoryear{Martin \& Roy}{1995}]{martin95} Martin, P., \& Roy, J.-R. 1995, ApJ, 445, 161 

\bibitem[\protect\citeauthoryear{Newman et al.}{2012}]{newman12} Newman, S.~F., Shapiro Griffin, K., Genzel, R. et al. 2012, ApJ, 752, 111

\bibitem[\protect\citeauthoryear{O'dell \& Casta\~neda}{1984}]{odell84} O'Dell, C. R., \& Casta\~neda, H.~O. 1984, ApJ, 283, 158

\bibitem[\protect\citeauthoryear{Oey \& Kennicutt}{1993}]{oey93} Oey, M.~S., \& Kennicutt, R.C. 1993, ApJ, 411, 137

\bibitem[\protect\citeauthoryear{Osterbrock  \& Ferland}{2006}]{osterbrock06} Osterbrock D.~E., Ferland G.~J., 2006, agna.book,  

\bibitem[\protect\citeauthoryear{Paturel et al.}{2003}]{paturel03} Paturel G., Petit C., Prugniel P., Theureau G., Rousseau J., Brouty M., Dubois P., Cambr{\'e}sy L., 2003, A\&A, 412, 45 

\bibitem[\protect\citeauthoryear{Puech et al.}{2006}]{puech06} Puech, M., Flores, H., Hammer, F. , Lehnert, M.~D., 2006, A\&A, 455, 131,134.

\bibitem[\protect\citeauthoryear{Pindao et al.}{2002}]{pindao02} Pindao, M., Schaerer, D., Gonz\'alez Delgado, R.~M.,  Stasi\'nska, G. 2002, A\&A, 394, 443

\bibitem[\protect\citeauthoryear{Ramsbottom et al.}{1996}]{Ramsbottom96} Ramsbottom, C. A. and Bell, K. L. and Stafford, R. P. 1996, Atomic Data and Nuclear Data Tables, 63, 57.

\bibitem[\protect\citeauthoryear{Rela{\~n}o et al.}{2010}]{relano10} Rela{\~n}o M., Monreal-Ibero A., 
V{\'{\i}}lchez J.~M., Kennicutt R.~C., 2010, MNRAS, 402, 1635

\bibitem[\protect\citeauthoryear{Rupke et al.}{2010}]{rupke10}  Rupke D.~S.~N.,  Kewley, L.~J., Chien, L.-H 2010, ApJ, 710, L156

\bibitem[\protect\citeauthoryear{Rich et al.}{2011}]{rich11}   Rich, J.~A., Kelley, L.~J., Dopita, M.~A. 2011, ApJ, 734, 87

\bibitem[\protect\citeauthoryear{Rich et al.}{2012}]{rich12}   Rich, J.~A., Torrey, T., Kelley, L.~J., Dopita, M.~A., Rupke, D.~S.~N. 2012, ApJ, 753, 5

\bibitem[\protect\citeauthoryear{Schaerer et al.}{2000}]{shaerer00} Schaerer, D., Guseva, N.~G., Izotov, Y.~I.,  Thuan, T.~X. 2000, A\&A, 362, 53

\bibitem[\protect\citeauthoryear{Sekiguchi \& Wolstencroft}{1993}]{sekiguchi93} Sekiguchi, K., \& Wolstencroft, R.~D.\ 1993, MNRAS, 263, 349 

\bibitem[\protect\citeauthoryear{Stanghellini  et al.}{1989}]{stanghellini89}  Stanghellini, L., \& Kaler, J.~B. 1989, ApJ, 343, 811

\bibitem[\protect\citeauthoryear{Soto  \& Martin}{2012}]{soto12} Soto, K.~T.,  Martin, C.~L.  2012,  ApJS, 203, 3

\bibitem[\protect\citeauthoryear{Soto  et al.}{2012}]{soto12a} Soto, K.~T.,  Martin, C.~L., Prescott, M.~K.~M.,  Armus, L.   2012,  ApJS, 757, 86

\bibitem[\protect\citeauthoryear{Scudder  et al.}{2012}]{scudder12} Scudder, J.~M., Ellison, S.~L., Torrey, P., Patton, D.~R.,  Trevor Mendel, J. 2012, astroph/1207.479

\bibitem[\protect\citeauthoryear{Storchi-Bergmann et al.}{2007}]{thaisa07} Storchi-Bergmann, T., Dors, O.~L.,  Riffel, R.~A. et al. 2007,  ApJ, 670, 959

\bibitem[\protect\citeauthoryear{Tubbs}{1982}]{tubbs82} Tubbs, A.~D. 1982, ApJ, 255, 458

\bibitem[\protect\citeauthoryear{Verner et al.}{1987}]{verner87} Verner, D. A. and Verner, E. M. and Ferland, G. J.1987, Atomic Data and Nuclear Data Tables, 1, 64.

\bibitem[\protect\citeauthoryear{Veilleux et al.}{2005}]{veilleux05} Veilleux, S., Cecil, G.,  Bland-Hawthorn, J. 2005, ARA\&A, 43, 76

\bibitem[\protect\citeauthoryear{Veilleux \& Osterbrock}{1987}]{veilleux87} Veilleux, S., \& Osterbrock, D.~E. 1987, ApJS, 63, 295

\bibitem[\protect\citeauthoryear{Weilbacher et al.}{2000}]{weilbacher00} Weilbacher, P.~M., Duc, P.-A., Fritze v.~Alvensleben, U., Martin, P., \& Fricke, K.~J.\ 2000, A\&A, 358, 819 

\bibitem[\protect\citeauthoryear{Zaritsky, Kennicutt, 
\& Huchra}{1994}]{zaritsky94} Zaritsky D., Kennicutt R.~C., Jr.,
  Huchra J.~P., 1994, ApJ, 420, 87

\end{thebibliography}
\end{document}